\documentclass[reprint,showpacs,superscriptaddress,english,twocolumn,notitlepage,longbibliography,nofootinbib]{revtex4-1}
\usepackage{times}
\usepackage{graphicx}
\usepackage{float}
\usepackage{latexsym,amsmath,amssymb,bm,euscript}
\usepackage{color}
\usepackage{subfigure}
\usepackage{epstopdf}
\usepackage[colorlinks=true,linkcolor=blue,citecolor=blue,urlcolor=blue]{hyperref}
\usepackage{hyperref}
\usepackage{ulem}
\usepackage{cleveref}

\usepackage{tikz}
\usetikzlibrary{decorations.pathreplacing}

\usepackage{ytableau}

\definecolor{colorhhy}{rgb}{0.9, 0.17, 0.31}
\newcommand{\hhy}[1]{\textcolor{black}{#1}}

\crefname{appendix}{Appendix}{Appendices}
\crefname{equation}{Eq.}{Eqs.}
\crefname{figure}{Fig.}{Figs.}
\crefname{table}{Table}{Tables}
\crefname{section}{Appendix}{Appendices}
\crefname{enumi}{Point}{Points}
\creflabelformat{appendix}{[#2#1#3]}
\crefmultiformat{figure}{Figs.~#2#1\xdef\mycreffirstarg{#1}#3}%
 { and~#2#1#3}{, #2#1#3}{ and~#2#1#3}
\newcommand{\be}[0]{\begin{equation}}
\newcommand{\ee}[0]{\end{equation}}
\newcommand{\hubbard}[0]{\mathcal{U}}

\def\ba#1\ea{\begin{align}#1\end{align}}  
\newcommand{\up}[0]{\uparrow}
\newcommand{\dn}[0]{\downarrow}
\newcommand{\bmat}[0]{\begin{bmatrix}}
\newcommand{\emat}[0]{\end{bmatrix}}

\def\kk{\mathbf{k}}
\def\qq{\mathbf{q}}

\def\RR{\mathbf{R}}
\def\rr{\mathbf{r}}
\def\xx{\mathbf{x}}

\newcommand{\citeSI}[1]{\cref{#1}}

\makeatletter
\let\oldcref\cref

\newcommand{\AddCrefMap}[2]{%
  \expandafter\def\csname crefmap@#1\endcsname{#2}%
}

\renewcommand{\cref}[1]{%
  \expandafter\ifx\csname crefmap@#1\endcsname\relax
    \oldcref{#1}%
  \else
    \csname crefmap@#1\endcsname
  \fi
}
\makeatother
\AddCrefMap{app:set_up_model}{Appendix~II~\cite{SM}}

\AddCrefMap{app:eff_act}{Appendix~III~\cite{SM}}

\AddCrefMap{sec:flat_band}{Appendix~V~\cite{SM}}
\AddCrefMap{sec:atomic}{Appendix~IV~\cite{SM}}
\AddCrefMap{sec:fm_afm}{Appendix~VI, VII~\cite{SM}}
\AddCrefMap{sec:toy_model}{Appendix~VIII~\cite{SM}}

\begin{document}

\title{Ferromagnetism vs. Antiferromagnetism in Narrow-Band Systems: Competition Between Quantum Geometry and Band Dispersion}

\author{Haoyu Hu}
\affiliation{Department of Physics, University of Science and Technology of China, Hefei, Anhui 230026, China}
\affiliation{Department of Physics, Princeton University, Princeton, NJ 08544, USA}

\author{Oskar Vafek}
% \affiliation{Department of Physics, Florida State University, Tallahassee, Florida 32306, USA}
% \affiliation{National High Magnetic Field Laboratory, Tallahassee, Florida, 32310, USA}
\affiliation{W. I. Fine Theoretical Physics Institute and School of Physics and Astronomy,
University of Minnesota, Minneapolis, Minnesota 55455, USA}

\author{Kristjan Haule}
\affiliation{Center for Materials Theory, Department of Physics and Astronomy, Rutgers University, Piscataway, NJ 08854, USA}

\author{B.~Andrei Bernevig}
% \email{bernevig@princeton.edu}
\affiliation{Department of Physics, Princeton University, Princeton, NJ 08544, USA}
\affiliation{Donostia International Physics Center (DIPC), Paseo Manuel de Lardizábal. 20018, San Sebastián, Spain}
\affiliation{IKERBASQUE, Basque Foundation for Science, 48013 Bilbao, Spain}
\begin{abstract}
   Magnetism in narrow-band systems arises from the interplay between electronic correlations, quantum geometry, and band dispersion. In particular, both ferro and anti-ferro magnets are known to occur as ground states of (different) models featuring narrow bands.
   This poses the question of which is favored and under what conditions. In this work, we present a unified theoretical framework to investigate spin physics within narrow bands. By deriving an effective spin model, we show that the non-atomic wavefunction (quantum geometry) of the narrow bands generally favors ferromagnetic ordering, while band dispersion promotes antiferromagnetic correlations. 
   We find that the competition between these effects gives rise to a tunable magnetic phase and rich spin phenomena.
   Our approach offers a systematic way to study the magnetic properties of narrow-band systems, integrating the roles of wave function, band structure, and correlation effects. 
\end{abstract}
\maketitle

{\it Introduction~~}
Spin physics~\cite{auerbach2012interacting} has long been a central theme in condensed matter research, giving rise to a wide range of magnetic phenomena, including spin liquids \cite{QSL_rev,savary2016quantum,takagi2019concept}, spiral orders\cite{nagamiya1968helical,kimura2007spiral,sosnowska1982spiral}, skyrmions\cite{nagaosa2013topological,tokura2020magnetic,fert2017magnetic}, and altermagnetism\cite{vsmejkal2022emerging}. 
In the single-orbital Hubbard model, strong correlation effects emerge when the on-site repulsion dominates over the kinetic energy set by hopping between localized Wannier orbitals. 
At half-filling with dominant nearest-neighbor hopping, this typically results in magnetic ordering via antiferromagnetic superexchange
interactions\cite{PhysRev.115.2,auerbach2012interacting,macdonald_fractu_1988}. 
However, studies of moir\'e\cite{bistritzer_moire_2011,CAO18a,YAN19,LU19,STE20,SAI20,DE21a,OH21,TIA23,DI22a,CAL22d,AND21,HAD20,BAL20,CAO18,KER19,XIE19,SHA19,JIA19,CHO19,POL19,YAN19,LU19,STE20,SAI20,SER20,CHE20b,WON20,CHO20,NUC20,CHO21,SAI21,LIU21c,PAR21c,WU21a,CAO21,DAS21,TSC21,PIE21,STE21,CHO21a,XIE21d,DAS22,NUC23,YU23c} and geometrically frustrated lattices\cite{PhysRevB.111.125163,PhysRevLett.128.087002,mielke1991ferromagnetic,mielke1991ferromagnetism,mielke1992exact,kang_topological_2020,sun_observation_2022,huang_non-fermi_2024} have revealed a distinct route to interaction-driven magnetism. 
These systems feature narrow bands with bandwidths much smaller than a typical hopping amplitude of an atomic orbital; their flatness is instead a result of subtle destructive interference of the kinetic energy matrix elements~\cite{Regnault2022,cualuguaru2022general,ma_topological_2021,jiang_fege_2025,leykam_artificial_2018,bergman_band_2008,chen_decoding_2023}. 
These narrow bands usually carry nontrivial quantum geometry and topological characteristics\cite{cualuguaru2022general,Regnault2022,BER21,SON19,yang_topological_2012,AHN19,LIU19,ZOU18,PO19,LIA20a,HEJ19a,PhysRevB.102.165148,chen2025quantumgeometricdipoletopologicalboost}, and their flatness makes them particularly susceptible to interaction effects\cite{PhysRevLett.129.047601,BER21a,BER21b,wu2020quantum,checkelsky_flat_2024,PhysRevLett.129.047601,deng_theory_2025,chen_emergent_2024,chen_metallic_2025,tian_evidence_2023,peri_fragile_2021,SHI22a,CAL23,TOM19,CAO20,ZON20,LIS21,BEN21,LIA21c,ROZ21,SAI21a,LU21,HES21,DIE23,HUB22,GHA22,JAO22,PAU22,GRO22,ZHO23a,PhysRevX.15.021087}. 
These systems do not fall within the conventional strong-coupling regime, defined by interactions dominating over bare hopping, thereby invalidating the applicability of the superexchange mechanism.
% Instead, it resides in a correlated regime beyond the reach of single-orbital Hubbard models.
Studies of such flat bands, especially in the idealized perfectly flat-band limit, have uncovered ferromagnetic ground states arising from the quantum geometry of the Bloch wavefunctions\cite{KAN19,BER21a,BER21b,PhysRevLett.128.087002,mielke1991ferromagnetic,mielke1991ferromagnetism,mielke1992exact,julku_geometric_2016,liang_band_2017,huhtinen_revisiting_2022,herzog-arbeitman_superfluid_2022,herzog-arbeitman_many-body_2022,peotta_superfluidity_2015,torma_superconductivity_2022,TSC21,PhysRevLett.123.237002,rossi2021quantum,PhysRevLett.132.036001,kitamura2025quantumgeometricferromagnetismsingular,chen2025quantumgeometricdipoletopologicalboost}.
These findings suggest a new class of magnetic phases governed not by real-space exchange interactions but by the geometric properties of the electronic bands\cite{torma2023essay,yu2025quantumgeometryquantummaterials}.

In realistic quantum materials, finite band dispersion and nontrivial quantum geometry typically coexist. Consequently, both conventional superexchange coupling and geometry-induced ferromagnetic correlation can jointly influence the magnetic ground state. 
However, a unified theoretical framework that simultaneously accounts for band dispersion, quantum geometry, and electronic interactions has been lacking. 
In this work, we bridge this gap by developing a comprehensive formalism that captures the interplay among these key ingredients shaping magnetism in narrow-band systems, and explains how and why certain narrow bands are ferromagnetic while others are antiferromagnetic. 
We investigate systems with half-filled narrow bands near the Fermi energy, characterized by both finite dispersion and nontrivial quantum geometry. 
We assume that the bandwidths of narrow bands are smaller than the interaction strength, placing the system in a strongly correlated regime. However, the bare hopping amplitude between atomic orbitals is not necessarily smaller than the interaction scale. 
Under these conditions, we derive an effective spin-spin interaction model that captures the magnetic behavior arising from the interplay of interactions, band dispersion, and quantum geometry. Our analysis shows that the quantum geometry tends to favor ferromagnetic order, while band dispersion promotes antiferromagnetic correlations. The competition between these effects potentially gives rise to a rich magnetic phase diagram. Within our effective model, we also estimate the transition point between ferromagnetic and antiferromagnetic phases, providing a microscopic perspective on interaction-driven magnetism in narrow-band systems.

Finally, we comment on the relation to, and distinction from, the early study in Ref.~\cite{PhysRev.115.2}. Ref.~\cite{PhysRev.115.2} showed that hopping between Wannier orbitals generates antiferromagnetic superexchange couplings, while the finite spatial extent of atomic Wannier orbitals gives rise to a ferromagnetic coupling. 
{
We emphasize that this ferromagnetic coupling already emerges at the level of model construction.  
Upon projecting the Coulomb repulsions onto a set of atomic Wannier orbitals, the ferromagnetic coupling appears as an interaction term in the resulting interacting tight-binding Hamiltonian, and becomes relatively weak for well-localized Wannier orbitals.
However, Ref.~\cite{PhysRev.115.2} did not fully address the role of the Bloch-wavefunction structure. This structure arises from the hybridization among Wannier orbitals when transforming from the Wannier-orbital basis to the band basis. Through this process, the Bloch bands can acquire nontrivial quantum geometry and even topological character.} 
Our study starts from {a tight-binding model with well-localized Wannier orbitals subject only to local Hubbard interactions.} 
\hhy{Therefore, due to the localized nature of the Wannier orbitals, the direct-exchange coupling discussed in Ref.~\cite{PhysRev.115.2} is already assumed to be negligible at the model level.} 
Nevertheless, our approach explicitly incorporates the contributions from the Bloch-wavefunction structure as well as the Bloch-band dispersion. 
We show that band dispersion promotes antiferromagnetism, whereas quantum geometry, reflecting the non-atomic nature of the Bloch wavefunctions, favors ferromagnetism. 
{In the atomic limit of a narrow band with well-localized Wannier orbitals, our approach reduces to the same antiferromagnetic correlations obtained in Ref.~\cite{PhysRev.115.2}. At the same time, our framework captures the additional contribution arising from the wavefunction structure of the Bloch bands, an effect absent in the earlier treatment.} 
Thus, our work is in the same spirit as Ref.~\cite{PhysRev.115.2}, but advances it by including wavefunction effects and providing a unified theory of magnetism in narrow-band systems.
\hhy{
Moreover, our framework can be applied to topologically nontrivial systems, where the narrow bands near the Fermi energy do not admit a single localized Wannier-orbital representation. Such systems are therefore also beyond the scope of Ref.~\cite{PhysRev.115.2}.
}

\begin{figure}
    \centering
    \includegraphics[width=1.0\linewidth]{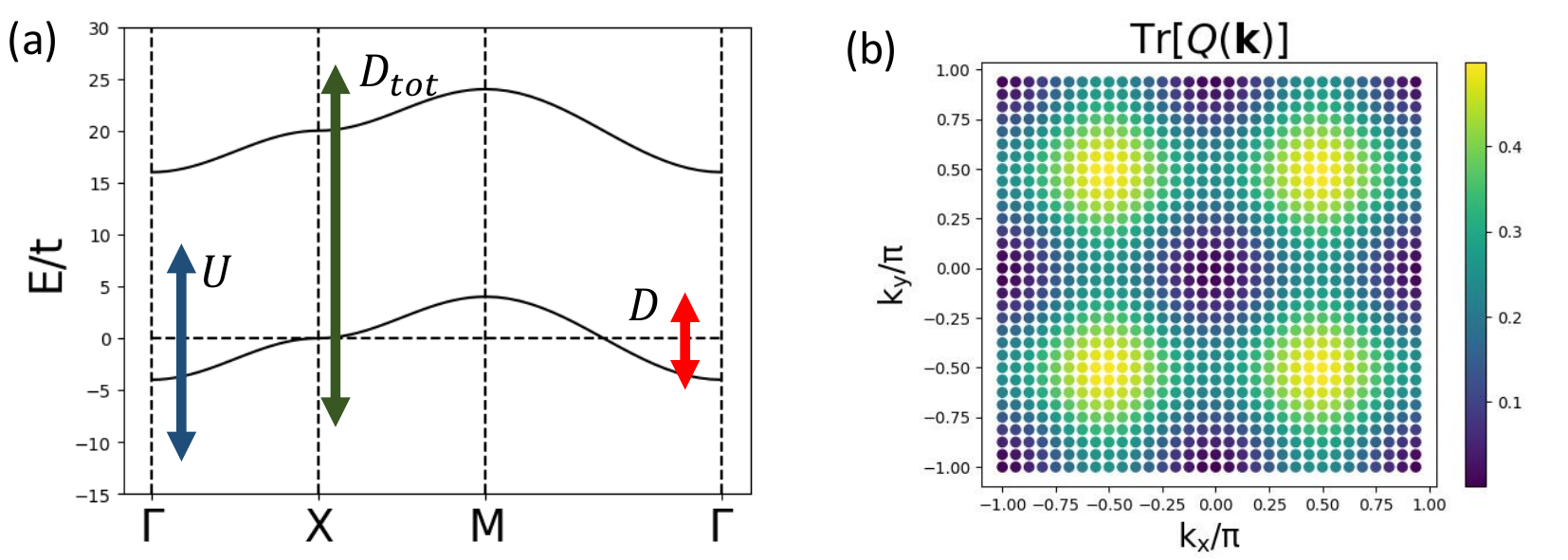}
    \caption{ Panel (a) illustrates a narrow-band system, where a narrow band with bandwidth $D$ emerges near the Fermi energy. The interaction strength $\hubbard $ is much larger than $D$, indicating strong correlations within the narrow band. Additional bands may also be present, resulting in a total electronic bandwidth $D_{\text{tot}}$ that exceeds $\hubbard$. 
    (b) Quantum geometry of the narrow band of the toy model defined in Eq.~\ref{eq:main:toy_model}. }\label{fig:main:narrow_band_sys}
\end{figure}

{\it Multi-orbital Hubbard model with narrow bands}~~
We start from a multi-orbital model with the following Hamiltonian 
\begin{align}
    &H = H_0 + H_U\nonumber\\ 
    &H_0 = \sum_{ab,\RR,\RR',\sigma} t_{\kk,ab}c_{\kk,a,\sigma}^\dag c_{\kk,b,\sigma} \nonumber\\ 
    &H_U = \sum_{\RR}\hubbard_a c_{\RR,a,\up}^\dag c_{\RR,a,\up} c_{\RR,a,\dn}^\dag c_{\RR,a,\dn} 
\end{align}
where $c_{\kk,a,\sigma}^\dag$ creates an electron with momentum $\kk$, flavor (sublattice) index $a=1,...,n_{sub}$ and spin $\sigma$. 
$t_{ab}(\kk)$ characterizes the kinetic term arising from short-range real-space hopping, and $\hubbard_a$ denotes the on-site Hubbard interaction. We note that, in real systems, additional interactions such as Hund’s coupling and density–density interactions between different flavors may arise. In this work, we focus on the case where electrons of different flavors are located at different positions within the unit cell, implying a weak Hund’s-type coupling, which typically occurs between electrons in different atomic orbitals but at the same position. As for the density–density interactions between electrons of different flavors, these terms can be treated at the Hartree–Fock level, absorbed into the kinetic term, and are less relevant to the magnetic correlations of interest here. 
We assume the system develops narrow bands near the Fermi energy, with a bandwidth smaller than the interaction scale characterized by $\hubbard = \mathrm{mean}_a,\hubbard_a$.  
% In addition, we consider the parameter region where the interaction scale, characterized by $\hubbard=\text{mean}_a\hubbard_a$, is much larger than the
% bandwidth of the narrow bands $D$ near the Fermi energy. 
Other bands beyond the narrow bands may also exist in the system, contributing to a total bandwidth $D_{{tot}}$ that can be much larger than $D$. However, we assume that the interaction scale $\hubbard $ is smaller than $D_{{tot}}$ (if such additional bands exist), so that only the narrow bands serve as the relevant low-energy degrees of freedom.  
An illustration of such narrow band systems has been shown in Fig.~\ref{fig:main:narrow_band_sys} (a).

{\it Non-interacting band structures}~~
The hopping matrix $t_{\kk,ab}$ in momentum space can be diagonalized into its eigenvalues $\epsilon_{\kk,n}$ and eigenvectors $U_{\kk,an}$, which represent the band dispersion and Bloch wavefunctions, respectively. Specifically, they satisfy the eigenvalue equation
\begin{equation}
    \sum_b t_{\kk,ab} \, U_{\kk,bn} = \epsilon_{\kk,n} \, U_{\kk,an}.
\end{equation}  
We focus on a set of narrow bands labeled by $n = 1, \dots, n_{\text{flat}}$ near the Fermi energy. The dispersion of these bands is written as $
\epsilon_{\kk,n} = \epsilon_0 + \delta \epsilon_{\kk,n}$
where $\epsilon_0 = \text{mean}_{\kk,\, n=1,\dots,n_{\text{flat}}} \, \epsilon_{\kk,n}$ denotes the average energy of the narrow bands, and $\delta \epsilon_{\kk,n}$ captures the deviation from this mean.
We assume that the average energy coincides with the Fermi level ($\epsilon_0 = 0$), {so that the narrow bands lie at the Fermi energy and are half-filled.} 
The orbital weight of $a$-th electrons in the narrow bands is defined as $
A_a = \frac{1}{N} \sum_{n=1}^{n_{\text{flat}}} \sum_{\kk} |U_{\kk,an}|^2$ 
where $N$ is the number of unit cells.  
In addition, we define the following two quantities to characterize the wavefunction structure and dispersion of the $n$-th narrow band
\begin{align}
    \label{eq:main:def_QM}
   &Q_{\mu\nu,n}=  \frac{1}{N}\sum_{\mu,\kk,ab}  
   \partial_{k^\mu}U_{\kk, an}^* 
   \bigg( \delta_{a,b} 
   - U_{\kk,an} U_{\kk,bn}^*\bigg)\partial_{k^\nu}U_{\kk,bn}  \nonumber\\ 
&M_{\mu\nu,n} = \frac{1}{N}\sum_{\kk} \partial_{k^\mu} \bigg(\delta \epsilon_{\kk,n} \bigg) 
\partial_{k^\nu} \bigg(\delta \epsilon_{\kk,n}\bigg) 
\end{align} 
$Q_{\mu\nu,n}$ is the quantum geometry\cite{torma2023essay} of the $n$-th band. $M_{\mu\nu,n}$ reflects the band dispersion. Since $\partial_{k^\mu}(\delta \epsilon_{\kk,n})$ measures the Fermi velocity, $M_{\mu\nu,n}$ quantifies the degree of band dispersiveness. As we will show later, these two quantities together can be used to characterize the magnetic correlations of the system.

{\it Effective spin model}~~
We now derive an effective theory to describe magnetic correlations in narrow-band systems. We perform a Hubbard–Stratonovich transformation, introducing an auxiliary bosonic field $\phi_{\RR,a}^{\mu}$ that characterizes the $\mu \in {x, y, z}$ component of the magnetic moment for electrons in unit cell $\RR$ with flavor index $a$ \cite{altland2010condensed,nagaosa1999quantum}. 
We separate the bosonic fields into two parts $\phi^\mu_{\RR,a} = \phi_{0,a_i}n^\mu_{\RR,a_i}$, 
where $\phi_{0,a_i}$ denotes the size of the local moment, and $n^\mu_{\RR,a_i}$ is 
a unit vector that describes the direction of the local moment and characterizes the magnetic order.

We integrate out our electron fields, which gives an effective action of the spin fields $\phi$ taking the form of 
$-\text{Tr}\log[G_{x_i,x_j}^{-1} + \frac{1}{2\beta}\sum_\mu \hubbard_{a_i}\phi_{x_i}^\mu \sigma^\mu \delta_{x_i,x_j}]$\cite{altland2010condensed,nagaosa1999quantum,PhysRevB.41.11457}, $x_{i/j} = (\RR_{i/j},a_{i/j})$ labels both the unit cell position and the flavor index. $G$ is the non-interacting Green's function of the electrons 
defined as 
\ba 
    G_{x_i,x_j}(\tau) = -\frac{1}{2}\sum_\sigma \langle T_\tau c_{x_i,\sigma}(\tau) c^\dag_{x_j,\sigma}(0)\rangle_{H_0}
\ea 
At this step, the action of the spin fields is exact without any approximations. 
To gain more insights on spin physics, we now simplify the action by making a ``gradient expansion''. To do so, we decompose the Green's function into local $G_{loc,a_i} (= G_{x_i,x_i})$ and non-local $G_{x_i,x_j}'$ part (see \citeSI{app:set_up_model})
\begin{align}
    G_{x_i,x_j}(\tau) = \delta_{x_i,x_j}G_{loc,a_i}(\tau) + (1-\delta_{x_i,x_j})G'_{x_i,x_j}(\tau)
\end{align}
The local component captures the on-site electronic properties, while the non-local component describes correlations between electrons of different $(\RR,a)$ indices. Therefore, the non-local component induces non-local magnetic correlations that govern the magnetic ordering of the system. 
We treat the non-local Green's function $G'$ as a perturbation and perform an expansion (see \citeSI{app:eff_act}). This approach is conceptually similar to a gradient expansion, where spatial correlations are systematically incorporated through gradient terms. 

At zeroth order in $G'$, the effective action is local (atomic) and gives 
% the effective action of $\phi_{0,a_i}$ fields takes the form of 
\begin{align}
    \label{eq:main:sphi0}
    S_{\phi_0} = \sum_a \bigg[ \frac{\hubbard_{a}}{4}\phi_{0,a}^2-\log(4\cosh(\hubbard_{a}A_a\phi_{0,a}/2) )\bigg] 
\end{align}
The saddle-point solution of Eq.~\ref{eq:main:sphi0} ($\frac{ \delta S_{\phi_0}}{\delta{\phi_0}}=0$) leads to a critical temperature $T_{c,a}=A_a^2\hubbard_a/4$. In the case where all flavors of electrons are equivalent with $\hubbard_a=\hubbard$ and $A_a=A$, we have $T_c= A^2\hubbard/4$. 
Below the critical temperature $T<T_c$, we have $\phi_{0,a}\ne 0 $, indicating the formation of local moments. 
In practice, $T_c$ can be understood as the temperature scale of the local-moment formation. 
At low-temperature limit $T\rightarrow 0$, electrons form local moments with size $\phi_{0,a}=A$. 
In addition, at zeroth-order, there is also a Berry phase term, $S_B$, that emerges for the spin fields $n^{\mu}_{\RR,a}$, taking the formula of the conventional Berry phase term of spin operators~\cite{auerbach2012interacting}.

The contribution from the first-order term in $G'$ always vanishes. The second-order term leads to an effective spin-spin interaction 
\begin{align}
    S_J = \int_\tau \sum_{x_i,x_j}J_{x_i,x_j}\bm{n}_{x_i} \cdot \bm{n}_{x_j}
\end{align} 
The effective spin-spin coupling 
is generated by the non-local Green's function $G'$ and takes the form of 
{\small 
\begin{align}
\label{eq:main:spin_spin_coupling_full}
&J_{x_i,x_j} \nonumber\\
= \frac{1}{2\beta} &\sum_{i\omega} \frac{\hubbard_{a_j}\hubbard_{a_i}A_{a_j}A_{a_i} G'_{x_i,x_j}(i\omega)G'_{x_j,x_j}(i\omega) e^{i\omega 0^+}/(2\beta)}{\bigg[1-\left(\frac{A_{a_i}^2\hubbard_{a_i}G_{loc,a_i}(i\omega)}{2i\omega}\right)^2\bigg]\bigg[1-\left(\frac{A^2_{a_j}\hubbard_{a_j}G_{loc,a_j}(i\omega)}{2i\omega}\right)^2\bigg]}
\end{align}
}
where $i\omega$ is the Matsubara frequency. 
For a given system, one could directly evaluate \cref{eq:main:spin_spin_coupling_full} to extract the spin-spin correlations. 

Higher-order terms in $G'$ give rise to multi-spin interactions of the form $n_{\RR,i}^\mu n_{\RR',j}^\nu n_{\RR'',n}^{\delta} \cdots$. These terms are expected to be less relevant due to their higher-order nature. 
% However, they can, in principle, be systematically evaluated. 
By combining the Berry phase term $S_B$ with the spin-spin interaction term $S_J$, we obtain an effective action that captures the magnetic correlations of the system.

To highlight the respective roles of band dispersion and wavefunction structure, we further simplify \cref{eq:main:spin_spin_coupling_full} by assuming that all electron flavors are equivalent, i.e., $\hubbard_a = \hubbard$ and $A_a = A$. Given that the narrow-band dispersion $\delta \epsilon_{\kk,n}$ is much smaller than the interaction strength $\hubbard $, we expand the expression in powers of $\delta \epsilon_{\kk,n}/\hubbard$. This yields the following effective spin-spin couplings (\citeSI{app:eff_act}):
\begin{widetext}
\begin{align}
    \label{eq:main:spin_spin_coupling}
   & J_{x_i,x_j} =J^1_{x_i,x_j} +J^2_{x_i,x_j}+J^3_{x_i,x_j} \nonumber\\ 
   & J^1_{x_i,x_j}= -\frac{\hubbard}{4} \bigg| A_{x_i,x_j} \bigg|^2 
   % A_{x_j,x_i} 
   ,\quad\quad  J^2_{x_i,x_j} = \frac{1}{A^4\hubbard}\bigg|B_{x_i,x_j}\bigg|^2 
    \nonumber\\ 
   &J^3_{x_i,x_j} =\frac{1}{2A^5\hubbard }
\bigg[
A\bigg( A_{x_i,x_j}C_{x_j,x_i} +A_{x_j,x_i}C_{x_i,x_j}\bigg) 
-3 B_{x_i,x_i} \bigg( A_{x_i,x_j}B_{x_j,x_i} +A_{x_j,x_i}B_{x_i,x_j} \bigg) 
- 3 \frac{AC_{x_i,x_i}-2B_{x_i,x_i}^2}{A} |A_{x_i,x_j}|^2
\bigg] \nonumber\\
&\quad A_{x_i,x_j} = \sum_{\substack{\kk\\n=1,...,n_{flat}} } U_{\kk,a_in}U_{\kk,a_jn}^*\frac{ e^{i\kk\cdot(\RR_i-\RR_j+\bm{r}_{a_i} -\bm{r}_{a_j})}}{N}
   ,\quad B_{x_i,x_j}=\sum_{\substack{\kk\\n=1,...,n_{flat}}} \delta \epsilon_{\kk,n}U_{\kk,a_in}U_{\kk,a_jn}^*\frac{ e^{i\kk\cdot(\RR_i-\RR_j+\bm{r}_{a_i} -\bm{r}_{a_j})}}{N}
   \nonumber \\ 
&\quad 
C_{x_i,x_j}=\sum_{\kk,n=1,...,n_{flat}} (\delta \epsilon_{\kk,n})^2U_{\kk,a_in}U_{\kk,a_jn}^*\frac{ e^{i\kk\cdot(\RR_i-\RR_j+\bm{r}_{a_i} -\bm{r}_{a_j})}}{N}
% \nonumber\\
% &\quad A =A_{x_i,x_i},\quad B= B_{x_i,x_i},\quad C= C_{x_i,x_i}
\end{align}
\end{widetext}
where we have dropped the high-order terms, which are at the order of $O( \hubbard(\frac{|\delta \epsilon_{\kk,n}|}{\hubbard})^3)$. 
Eq.~\ref{eq:main:spin_spin_coupling} is the main result of this work.

{\it Magnetic correlations}~~ 
We now analyze each term in Eq.~\ref{eq:main:spin_spin_coupling} individually, where the spin-spin couplings are decomposed into three distinct contributions. The first term, $J^1_{x_i,x_j}$, is non-positive and therefore favors ferromagnetic order. Importantly, this term depends solely on the wavefunction structure of the narrow bands. 
In the limit of a perfectly flat band (i.e., $\delta \epsilon_{\kk,n} = 0$) with finite quantum geometry, which reflects both the momentum-space variation of the wavefunction and its non-atomic nature, $J^1_{x_i,x_j}$ is the only non-vanishing contribution (\citeSI{sec:flat_band}).
Consequently, the system develops a ferromagnetic ground state at low temperatures, {consistent with previous studies of flat-band ferromagnetism in the half-filled case}\cite{KAN19,BER21a,BER21b,PhysRevLett.128.087002,mielke1991ferromagnetic,mielke1991ferromagnetism,mielke1992exact,ABO20,julku_geometric_2016,liang_band_2017,huhtinen_revisiting_2022,herzog-arbeitman_superfluid_2022,herzog-arbeitman_many-body_2022}. 
\hhy{A concrete example of quantum-geometry-induced ferromagnetism in a topological flat band is further discussed in \cref{sec:flat_band}.}
The second term, $J^2_{x_i,x_j}$, is always non-negative and thus favors antiferromagnetic ordering. In the case of a system with a single atomic orbital and a trivial wavefunction, i.e., $U_{\kk,(a=1,n=1)} = 1$, and finite bandwidth, this term reduces to the conventional superexchange interaction proportional to $t^2/U$, where $t$ denotes the hopping amplitude (\citeSI{sec:atomic}). However, in a more generic multiband system, this term is related to the band dispersion rather than to the bare hopping amplitudes themselves. 
The third term, $J^3_{x_i,x_j}$, depends on both the band dispersion and wavefunction structure. Its sign and magnitude are generally non-universal and can favor either ferromagnetism or antiferromagnetism, depending on the specific parameters. 
In general, we conclude that the non-atomic wavefunction
of the narrow bands tends 
to stabilize ferromagnetism, while the dispersion of the system tends to destroy the ferromagnetism by enhancing the antiferromagnetic correlation.  For a generic system with both a non-atomic wavefunction and finite dispersion, the competition among $J^{1}_{x_i,x_j}$, $J^{2}_{x_i,x_j}$, and $J^{3}_{x_i,x_j}$ may lead to magnetic frustration and stabilize exotic magnetic orders.

{\it Competition between quantum geometry and band dispersion}~~ 
To analyze the competing effect between quantum geometry and the dispersion of the system, we take our effective spin model characterized by spin-spin coupling $J_{x_i,x_j}$, and obtain the condition where the ferromagnetic state is no longer favored. 
We notice that, by treating the spin as classical spin, the energy of a given spin configuration $\{\bm{n}_{x_i}\}_{x_i}$ is 
\begin{align}
    &E/N = \sum_\qq J_{\qq, ab}\bm{n}_{-\qq,a}\cdot \bm{n}_{\qq,b}
\end{align}
where 
\begin{align}
&J_{\qq,ab} = \frac{1}{N}\sum_{\RR,\RR'}
    J_{(\RR,a),(\RR',b)}e^{i\qq\cdot(\RR'+\rr_b-\RR-\rr_a)}\nonumber\\ 
 &
    n_{\qq,a} = \frac{1}{N}\sum_{\RR}\bm{n}_{x= (\RR,a)}e^{-i\qq\cdot(\RR+\rr_a)}
\end{align}
We denote by $E_{\qq, \mathrm{lowest}}$ the lowest eigenvalue of $J_{\qq, ab}$ at each momentum $\qq$. The magnetic wavevector is then determined by the value of $\qq$ that minimizes $E_{\qq, \mathrm{lowest}}$. 
To explore the competition between dispersion and quantum geometry, we begin from the perfect-flat-band limit ($\delta \epsilon_{\kk,n}=0$), where a ferromagnetic state with $\qq = 0$ is stabilized. We then gradually introduce dispersion and examine the stability of this ferromagnetic state by evaluating the Hessian matrix of $E_{\qq, \mathrm{lowest}}$ at $\qq = 0$:
\begin{align}
    H_{\mu\nu} = &\frac{1}{2} \partial_{q_\mu}\partial_{q_\nu} E_{\qq,lowest}\bigg|_{\qq=0}
\end{align}
If the Hessian matrix $H_{\mu\nu}$ possesses negative eigenvalues, it indicates that $\qq = 0$ no longer minimizes $E_{\qq, {lowest}}$, signaling an instability of the ferromagnetic state toward a nonzero-$\qq$ (antiferromagnetic) ordering.

To simplify the analysis, we again consider a system with a single flat band near the Fermi energy and assume all electron flavors are equivalent, i.e., $\hubbard_a = \hubbard$ and $A_a = A$. Under this condition, $H_{\mu\nu}$ takes the form of (\citeSI{sec:fm_afm})
\begin{equation}
\label{eq:Hessian}
H_{\mu\nu} = A\hubbard \left[ \frac{Q_{\mu\nu,n=1}}{8} - \frac{M_{\mu\nu,n=1}}{4A^4\hubbard^2} \right],
\end{equation}
where $Q_{\mu\nu,n}$ and $M_{\mu\nu,n}$ (Eq.~\ref{eq:main:def_QM}) characterize the quantum geometry and band dispersion, respectively. 
The stability of the ferromagnetic state is determined by the eigenvalues of $H_{\mu\nu}$. When the smallest eigenvalue of $H_{\mu\nu}$ becomes negative, the system becomes unstable toward magnetic ordering with a finite wavevector $\qq \ne 0$. 
We note that both $Q_{\mu\nu,n=1}$ and $M_{\mu\nu,n=1}$ are positive semi-definite matrices. Therefore, when the quantum geometric contribution dominates, the ferromagnetic state remains stable. In contrast, if the dispersion term outweighs the geometric contribution, the ferromagnetic state becomes unstable and the system tends to develop magnetic order at a finite wavevector.

\begin{figure}
    \centering
    \includegraphics[width=0.7\linewidth]{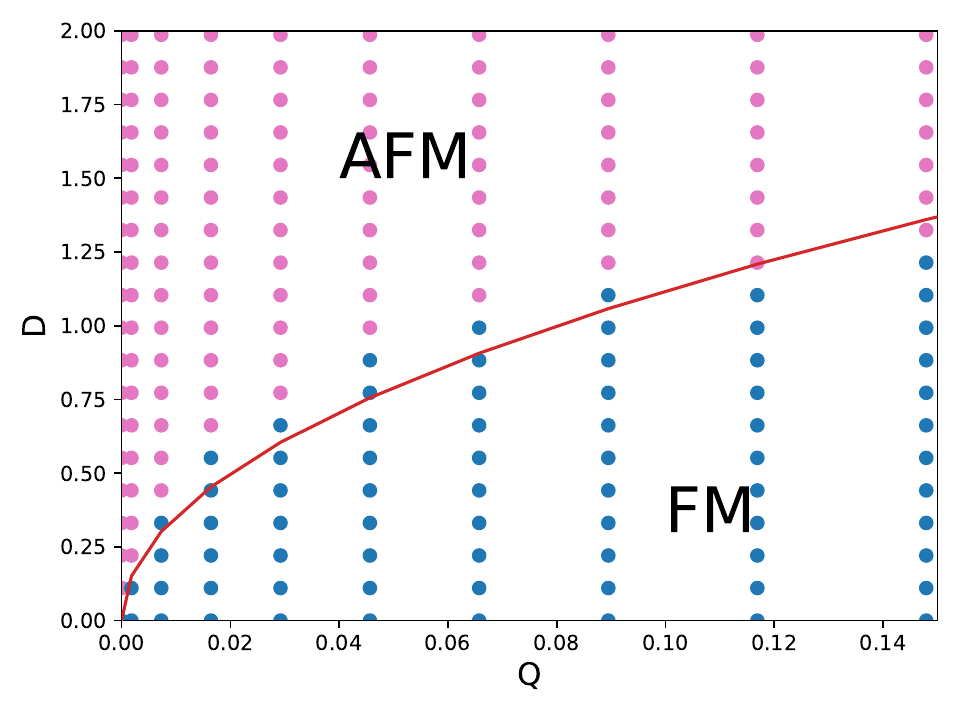}
    \caption{Hartree Fock phase diagram of the toy model defined in Eq.~\ref{eq:main:toy_model}~\cite{data_zenodo_20484319}.
    Pink and blue characterize two different ground states from the Hartree-Fock calculations, where pink denotes the antiferromagnetic phase and blue denotes the ferromagnetic phase. 
    $Q =\sum_{\mu}Q_{\mu\mu,n=1}$ denotes the quantum geometry and $D = 8t$ denotes the bandwidth of the narrow band. 
    The red curve denotes the phase boundary we obtained via the analytical expressions (Eq.~\ref{eq:Hessian}). }
    \label{fig:main:pd_toy_model}
\end{figure}

{\it Toy model}~~
To further investigate the competition between quantum geometry and band dispersion, we consider the following toy model defined on a bilayer square lattice, as introduced in Refs.~\cite{Hofmann2022,PhysRevLett.130.226001} 
\ba 
\label{eq:main:toy_model}
&H_0 = \sum_{s,\kk}
\begin{bmatrix}
    c_{\kk,+,\sigma}^\dag & c_{\kk,-,\sigma}^\dag 
\end{bmatrix}
\cdot 
\begin{bmatrix}
    \epsilon_\kk -\mu & v e^{i\alpha_\kk } \\
 v e^{-i\alpha_\kk } & \epsilon_\kk-\mu  
\end{bmatrix}\cdot 
\begin{bmatrix}
    c_{\kk,+,\sigma} \\ c_{\kk,-,\sigma}
\end{bmatrix}\nonumber
\\ 
& \epsilon_{\kk,1} = -2t (\cos(k_x) + \cos(k_y))\nonumber\\
&\alpha_\kk = \zeta (\cos(k_x)+\cos(k_y))
\ea 
with $\epsilon_\kk = -2t (\cos(k_x) + \cos(k_y)),\alpha_\kk = \zeta (\cos(k_x)+\cos(k_y))$. $c_{\kk,l,\sigma}^\dag$ creates an electron in layer $l = \pm$ with spin $\sigma$ and momentum $\kk$. 
We take the limit of $v =-\mu \gg U \gg t >0$. The bandwidth of narrowband is $D=8t$.
The gap between the narrow band near the Fermi energy and the remote band (high energy band) is $|v|$. 
The narrow band near the Fermi energy has $M_{\mu\nu,n=1}=\delta_{\mu,\nu}2t^2$ and quantum geometry $Q_{\mu\nu,n=1}= \delta_{\mu,\nu}\zeta^2/8$ (Eq.~\ref{eq:main:def_QM}). 
We can tune the dispersion and quantum geometry of the system individually by tuning $t$ and $\zeta$. 
A typical band structure has been shown in Fig.~\ref{fig:main:narrow_band_sys} ($t=1, \zeta=1,v=-\mu=10$).  
 We perform Hartree-Fock calculation for the electronic model $H_0+H_U$ with $\hubbard_1=\hubbard_2=\hubbard$. 
 We explore the phase diagram by independently tuning the bandwidth of the flat band ($D$) and the quantum geometric contribution $Q = \sum_\mu Q_{\mu\mu,n=1}$. The resulting Hartree-Fock phase diagram is shown in Fig.~\ref{fig:main:pd_toy_model} (\citeSI{sec:toy_model}).
We observe a quantum phase transition between antiferromagnetic and ferromagnetic phases. The phase boundary predicted by the Hessian analysis (\citeSI{eq:Hessian}) is given by $Q = 2D^2/\hubbard^2$, and is shown as the red curve in Fig.~\ref{fig:main:pd_toy_model}. Notably, this analytical prediction agrees well with the numerically determined phase boundary.

{\it Summary}~~
In this work, we have investigated the magnetic properties of narrow-band systems by developing a unified theoretical framework that incorporates the effects of both band dispersion and quantum geometry. {We focus on the regime of SU(2)-symmetric, half-filled narrow bands subject to Hubbard interactions, with the interaction strength exceeding the bandwidth of the narrow bands. By expanding in powers of the nonlocal Green’s function and truncating the result to two-body spin-spin interactions,} 
we derive an effective spin model that captures the essential magnetic correlations driven by the interplay of quantum geometry, band dispersion, and Hubbard interactions.

Our analysis uncovers a fundamental competition: the quantum geometry of the bands tends to favor ferromagnetic ordering, whereas the band dispersion promotes antiferromagnetic correlations. This interplay determines the magnetic ground state and gives rise to a rich phase diagram. Importantly, we derive analytical expressions for the spin-spin interactions in terms of the Bloch wavefunctions and band dispersions, elucidating how both features contribute to the effective magnetic coupling. In summary, we have developed a general and physically transparent framework for magnetism in narrow-band systems, identifying quantum geometry and band dispersion as competing mechanisms governing magnetic correlations. This work provides a foundation for understanding magnetic behavior in quantum materials from microscopic considerations.

\begingroup
  \renewcommand{\addcontentsline}[3]{}
  \begin{acknowledgments}
  B.A.B. and H.H. were supported by the Gordon and Betty Moore Foundation through Grant No. GBMF8685 towards the Princeton theory program, the Gordon and Betty Moore Foundation’s EPiQS Initiative (Grant No. GBMF11070), the Office of Naval Research (ONR Grant No. N00014-20-1-2303), the Global Collaborative Network Grant at Princeton University, the Simons Investigator Grant No. 404513, the Princeton Global Network, the NSF-MERSEC (Grant No. MERSEC DMR 2011750), the Simons Collaboration on New Frontiers in Superconductivity (Grant No. SFI-MPS-NFS-00006741-01 and SFI-MPS-NFS-00006741-06), and the Schmidt Foundation at the Princeton University. 
  H.H. acknowledges support from European Research Council (ERC) under the European Union’s Horizon 2020 research and innovation program (Grant Agreement No. 101020833), which supported the initial stage of this work. 
  \end{acknowledgments}
\endgroup

\clearpage 
\newpage 
\onecolumngrid
\begin{center}
\textbf{Supplementary Materials}\\ 
% \textbf{Haoyu Hu}
\end{center}
%%%%%%%%%% Merge with supplemental materials %%%%%%%%%%
%%%%%%%%%% Prefix a "S" to all equations, figures, tables and reset the counter %%%%%%%%%%

\renewcommand{\thefigure}{S\arabic{figure}}

\renewcommand{\thetable}{S\arabic{table}}

\renewcommand{\thesection}{S\arabic{section}}

\renewcommand{\theequation}{S\arabic{equation}}

\tableofcontents

\clearpage

\section{Model and effective action}
\label{app:spin_spin_int}
We consider the following multi-orbital system with Hubbard interactions
\ba 
S = \int_\tau \sum_{a,b,\RR,\RR',\sigma} c_{\RR,a,\sigma}^\dag(\tau) (\partial_\tau \delta_{a,b}\delta_{\RR,\RR'} + t_{ab}(\RR'-\RR))c_{\RR',b,\sigma}(\tau-0^+) + \sum_{\RR,a} \hubbard_ac_{\RR,a,\up }^\dag(\tau) c_{\RR,a,\up}(\tau) c_{\RR,a,\dn}^\dag (\tau) c_{\RR,a,\dn} (\tau)
\ea 
where $t_{ij}(\RR'-\RR)$ is the hopping matrix, $\hubbard_a$ is the on-site Coulomb repulsion, and $a$ denotes the sublattice, orbital indices.
We first perform a Hubbard-Startonovich (HS) transformation to obtain the effective spin fields. For a given unit vector $\bm{n}_{\RR,i}$ ($|\bm{n}_{\RR,i}|=1$), the Hubbard interaction can be written as 
{
\ba 
\hubbard_{a}n_{\RR,a,\up} n_{\RR,a,\dn} =\frac{\hubbard_a}{4}(N^{charge}_{\RR,a})^2 - \frac{\hubbard_a}{4}\bigg(\sum_{\sigma',\sigma''}c^\dagger_{\RR,a,\sigma'}\bigg(\bm{\sigma}_{\sigma',\sigma''}\cdot \bm{n}_{\RR,a}\bigg) c_{\RR,a,\sigma''}\bigg)^2 
\label{eq:hubbard_u_sep}
\ea 
where $N^{charge}_{\RR,a} =\sum_\sigma c_{\RR,a,\sigma}^\dag c_{\RR,a,\sigma} $ is the charge operator. From the Gaussian integral, we obtain the following Hubbard-Stratonovich (HS) decoupling
% \textcolor{red}
{
\ba 
&\int D[\phi^c] e^{- \int_\tau [\frac{\hubbard_a}{4} [\phi^{c}_{\RR,a}(\tau)]^2 + i\frac{\hubbard_a}{2} \phi^{c}_{\RR,a}(\tau)N^{charge}_{\RR,a}(\tau)] } \propto  e^{- \frac{\hubbard_a}{4}\int_\tau [N^{charge}_{\RR,a}(\tau)]^2  } 
\nonumber\\ 
&\int D[\phi^s] e^{- \int_\tau \bigg[\frac{\hubbard_a}{4} [\phi^{s}_{\RR,a}(\tau)]^2 - \frac{\hubbard_a}{2} \phi^{s}_{\RR,a}(\tau) \bigg( 
\sum_{\sigma',\sigma''}c_{\RR,a,\sigma'}(\tau) \bm{\sigma}_{\sigma',\sigma''}\cdot \bm{n}_{\RR,a}c_{\RR,a,\sigma''}(\tau) \bigg) 
\bigg] } \nonumber\\ 
\propto&  e^{+\int_\tau \frac{\hubbard_a}{4} \bigg( 
\sum_{\sigma',\sigma''}c_{\RR,a,\sigma'}(\tau) \bm{\sigma}_{\sigma',\sigma''}(\tau)\cdot \bm{n}_{\RR,a}(\tau)c_{\RR,a,\sigma''}(\tau)
\bigg)^2  }
\ea 
} 
where $\phi^c$ and $\phi^s$ are the bosonic fields introduced via HS decoupling. We can observe that, integrating over $\phi^c$ and $\phi^s$ fields yields the first and second terms of \cref{eq:hubbard_u_sep} respectively.

Since the unit vector $\bm{n}_{\RR,a}$ is arbitrary, we take an additional integral over unit vector $\bm{n}_{\RR,a}$
% \ba 
% &\int D[\phi_s,\bm{n}] e^{- \int_\tau \bigg[\frac{\hubbard_a}{4} \phi^{s,2}_{\RR,a}(\tau) - \frac{\hubbard_a}{2} \phi^{s,2}_{\RR,a}(\tau) \bigg( 
% \sum_{\sigma',\sigma''}c_{\RR,a,\sigma'}(\tau) \bm{\sigma}_{\sigma,\sigma''}\cdot \bm{n}_{\RR,i}c_{\RR,a,\sigma''}(\tau) \bigg) 
% \bigg] }  \nonumber\\ 
% \propto & e^{+ \int_\tau \frac{\hubbard_a}{4} \bigg( 
% \sum_{\sigma',\sigma''}c_{\RR,a,\sigma'}(\tau) \bm{\sigma}_{\sigma,\sigma''}(\tau)\cdot \bm{n}_{\RR,a}(\tau)c_{\RR,a,\sigma''}(\tau)
% \bigg)^2  }
% \ea 
% \textcolor{red}
{\ba 
&\int D[\phi_s,\bm{n}] e^{- \int_\tau \bigg[\frac{\hubbard_a}{4} [\phi^{s}_{\RR,a}(\tau)]^2 - \frac{\hubbard_a}{2} \phi^{s}_{\RR,a}(\tau) \bigg( 
\sum_{\sigma',\sigma''}c_{\RR,a,\sigma'}(\tau) \bm{\sigma}_{\sigma',\sigma''}\cdot \bm{n}_{\RR,a}c_{\RR,a,\sigma''}(\tau) \bigg) 
\bigg] }  \nonumber\\ 
\propto & e^{- \int_\tau \frac{\hubbard_a}{2} c_{\RR,a,\up}^\dag c_{\RR,a,\up}c_{\RR,a,\dn}^\dag c_{\RR,a,\dn}
+ \frac{\hubbard_a}{4}\sum_{\sigma}c_{\RR,a,\sigma}^\dag c_{\RR,a,\sigma}
}
\ea 
}
where we note that 
\ba 
\frac{\hubbard_a}{2} c_{\RR,a,\up}^\dag c_{\RR,a,\up}c_{\RR,a,\dn}^\dag c_{\RR,a,\dn}
- \frac{\hubbard_a}{4}\sum_{\sigma}c_{\RR,a,\sigma}^\dag c_{\RR,a,\sigma} = - \frac{\hubbard_a}{4}
\bigg( 
c_{\RR,a,\up}^\dag c_{\RR,a,\up} - 
c_{\RR,a,\dn}^\dag c_{\RR,a,\dn} 
\bigg) ^2 
\ea 
still reproduces the second term of \cref{eq:hubbard_u_sep}.  This can be seen by taking $\bm{n}_{\RR,a} = (0,0,1)$ in \cref{eq:hubbard_u_sep}. 
This allows us to define a new vector field to simplify the notation
\ba 
\bm{\phi}_{\RR,a}(\tau) = \phi^s_{\RR,a}(\tau) \bm{n}_{\RR,a}(\tau)
\label{eq:spin_field_decomp}
\ea

% We can perform Hubbard-Stronnovich decoupling which gives
% \ba 
% \hubbard_a(\Delta_c^2 +\Delta_s^2) 
% -\hubbard c_{\RR,a}^\dag (i\Delta_c +\Delta_{s}\bm{\Omega}\cdot \bm{\sigma})c_{\RR,a}
% \ea 
% We can let 
% \ba 
% \bm{\phi}_{\RR,a} = \bm{\Omega}\Delta_s
% \ea 
The action of the system now becomes 
\ba 
S =& \int_\tau\bigg\{  \sum_{\RR,a}\frac{\hubbard_a}{4}\bigg[ [\phi^c_{\RR,a}(\tau)]^2 +[\bm{\phi}_{\RR,a}(\tau)]^2\bigg]\nonumber\\ 
&
+\sum_{\RR,a}\sum_{\sigma',\sigma''}c_{\RR,a,\sigma'}^\dag(\tau) \bigg(\partial_\tau +i\frac{\hubbard_a}{2}\phi^c_{\RR,a}(\tau) \delta_{\sigma',\sigma''}-\frac{\hubbard_a}{2}\bm{\phi}_{\RR,a}(\tau)\cdot \bm{\sigma}_{\sigma',\sigma''}\bigg) c_{\RR,a,\sigma''}(\tau -0^+) + H_0(\tau)
\bigg\} 
\ea 
where $H_0(\tau) = \sum_{\RR,\RR',ab,\sigma}c_{\RR,a,\sigma}(\tau)t_{ab}(\RR'-\RR) c_{\RR',b,\sigma}(\tau)$ denotes the hopping term. 
The additional $-0^+$ ensures that $c^\dag_{\RR,a,\sigma'}(\tau)$ appears before $c_{\RR,a,\sigma''}(\tau-0^+)$ in the path-integral formula. 
Now we briefly discuss the physical meaning of the above action. $\phi_{\RR,a}^c$ describes the charge fluctuations of the system, and $\bm{\phi}_{\RR,a}$ describes the spin fluctuations of the system, where $\phi_{\RR,a}^s$ corresponds to the size of local moments and $\bm{n}_{\RR,a}$ describes the directions of the local moments. We can also observe their physical meaning from the saddle-point equations of $\phi^c_{\RR,a}$ and $\bm{\phi}_{\RR,a}$ fields which are
\ba 
&0 = \frac{\delta S}{\delta \bm{\phi}^{s}_{\RR,a}(\tau) }  = \frac{\hubbard_a}{2}\bm{\phi}_{\RR,a}(\tau)  - \frac{\hubbard_a}{2}  \sum_{\sigma',\sigma''} c_{\RR,a,\sigma'}^\dag(\tau) \bm{\sigma}_{\sigma',\sigma''}c_{\RR,a,\sigma''}^\dag(\tau)  \nonumber\\ 
\rightarrow &\bm{\phi}_{\RR,a}(\tau) 
= \sum_{\sigma',\sigma''} c_{\RR,a,\sigma'}^\dag(\tau) \bm{\sigma}_{\sigma',\sigma''}c_{\RR,a,\sigma''}^\dag(\tau)
\ea 
% which indicates $\phi_{\RR,i}^s$ describes the spin fluctuations of the electrons. As for the charge sector, the equation of motion of the $\phi^c$ field is 
and
\ba 
&0 = \frac{\delta S}{\delta \phi^c_{\RR,a}(\tau) } 
= \frac{\hubbard_a}{2}\phi_{\RR,i}^c(\tau) + i\frac{\hubbard_a}{2} \sum_\sigma c_{\RR,a,\sigma}^\dag c_{\RR,a,\sigma}(\tau) \nonumber\\ 
\rightarrow &\phi^c_{\RR,a} = -i\sum_\sigma c_{\RR,a,\sigma}^\dag c_{\RR,a,\sigma}
\label{eq:eom_charge}
\ea 

Since we are mostly interested in the fluctuations in the spin sectors, we could take the saddle-point approximation in the charge sectors by letting (which is obtained from Eq.~\ref{eq:eom_charge} by taking the expectation value of electron bilinear operators)
\ba 
\phi_{\RR,a}^c = -i\sum_\sigma \langle c_{\RR,a,\sigma}^\dag c_{\RR,a,\sigma}\rangle 
\ea 
By replacing $\phi_{\RR,a}^c$ with $-i\sum_\sigma \langle c_{\RR,a,\sigma}^\dag c_{\RR,a,\sigma}\rangle $, the charge field gives an on-site potential term and can be absorbed by $H_0$.
In addition, since our focus is on magnetic properties, we assume that no charge density wave develops, and thus take $\phi_{\RR,a}^c$ to be independent of $\RR$. 
For what follows, we omit $\phi_{\RR,a}^c$ in the action to simplify the notation. In general, we work in the large $\hubbard$ situation, when we expect the charge fluctuations to be suppressed near the half-filling of the flat band. 
%We leave a more comprehensive analysis, incorporating both spin and charge fluctuations across all fillings, for future investigation.

% \textcolor{red}{It seems very reasonable to freeze charge fluctuations by taking the saddle point. But we should nevertheless check the lowest order correction if we include them. Charge susceptibility is expected to be very small, hence I understand that this is reasonable.}

We now aim to derive an effective theory of the spin fields $\bm{\phi}_{\RR,a}$ based on the following action
\ba 
S = \int_\tau \sum_{\RR,a}\bigg[ \frac{\hubbard_a}{4}[\bm{\phi}_{\RR,a}(\tau)]^2
+\sum_{\sigma',\sigma''}c_{\RR,a,\sigma'}^\dag(\tau) \bigg(\partial_\tau  \delta_{\sigma',\sigma''}-\frac{\hubbard_a}{2}\bm{\phi}_{\RR,a}(\tau)\cdot \bm{\sigma}_{\sigma',\sigma''}\bigg) c_{\RR,a,\sigma''}(\tau)\bigg] +\int_\tau  H_0(\tau)
\ea 

We first separate the action into two parts
\ba 
S = &S_\phi + S_f \nonumber\\ 
S_\phi =& \int_\tau \sum_{\RR,a}\bigg[ \frac{\hubbard_a}{4}[\bm{\phi}_{\RR,a}(\tau)]^2\bigg] \nonumber\\ 
S_f = &\int_\tau \sum_{\RR,a}\bigg[ \sum_{\sigma}c_{\RR,a,\sigma}^\dag(\tau) \partial_\tau  c_{\RR,a,\sigma}(\tau-0^+)\bigg] + H_0(\tau) \nonumber\\ 
&+ \int_\tau \sum_{\RR,a,\sigma',\sigma''} c_{\RR,a,\sigma'}^\dag(\tau) \bigg(  -\frac{\hubbard_a}{2}\bm{\phi}_{\RR,a}(\tau)\cdot \bm{\sigma}_{\sigma',\sigma''}\bigg) c_{\RR,a,\sigma''}(\tau-0^+) 
\label{eq:action}
\ea

To simplify the notation, we introduce $x_i = (\RR_i,a_i)$ which denotes both the position of the electrons (including the unit cell position $\RR$ and the sublattice position $\rr_{a_i}$).
In addition, we let
\ba 
c_{x_i,\sigma}(\tau) = c_{\RR_i,a_i,\sigma}(\tau) ,\quad \bm{\phi}_{x_i}(\tau) = \bm{\phi}_{\RR_i,a_i}(\tau),\quad t_{x_j,x_i} = t_{a_j,a_i}(\RR_j-\RR_i)
\ea  
and introduce the operators in Matsubara frequency
\ba 
&c_{x_i,\sigma}(i\omega) =\int_0^\beta  c_{x_i,\sigma}(\tau) e^{i\omega \tau} d\tau  \nonumber\\ 
&\bm{\phi}_{x_i}(i\Omega) = \int_0^\beta  \bm{\phi}_{x_i}(\tau) e^{i\Omega \tau} d\tau 
\ea 
The action now can be written as 
\ba 
S =&S_\phi \nonumber\\ 
&+\frac{1}{\beta}\sum_{i\omega,i\omega',\xx_i,\xx_j,\sigma',\sigma''}
c_{x_i,\sigma}^\dag (i\omega) 
\left[ \left[ -i\omega
+ t_{x_i,x_j}\right] 
\delta_{\omega,\omega'}  \delta_{\sigma,\sigma'}
- \frac{\hubbard_{a_i}}{2\beta } \bm{\phi}_{x_i}(i\omega-i\omega')\cdot\bm{\sigma}_{\sigma,\sigma'} 
\right]e^{i\omega'0^+}c_{x_j,\sigma'} (i\omega') 
\label{eq:action_mats}
\ea 
We mention that $\bm{\phi}_{x_i}(i\omega-i\omega')$ indicates the $\bm{\phi}$ field also depends on the frequency $i\omega-i\omega'$.  
{
We have also incorporated the additional single-particle term $\hubbard_a/2c_{\RR,a,\sigma}^\dag c_{\RR,a,\sigma}$, arising from the Hubbard-Stratonovich transformation of the charge channel, into the hopping matrix $t_{x_i,x_j}$}. The additional $e^{i\omega' 0^+}$ factor ensures that the operator $c_{x_i,\sigma}^\dag$ appears before $c_{x_j,\sigma'}$ when transforming back to the imaginary-time domain. 
% $t_{x_i,x_j}$ are not anymore equal to those introduced above, because we have absorbed static charge part, i.e., $t^{new}_{x_i,x_j}=t_{x_i,x_j}+\frac{\hubbard}{2}\langle{n_{xi}}\rangle\delta_{xi,xj}$ }

We can use the following Gaussian integral of the Grassmann fields
\ba 
\int D[c, c^\dag] e^{ \sum_{ij} c_i^\dag M_{ij} c_j } = \text{det}[-M] = e^{\log(\text{det}[-M] ) } = 
e^{\text{Tr}[\log(-M)] } \propto e^{\text{Tr}[\log(M)] }
\label{eq:Grassmann_integral}
\ea 
We define the matrix
\ba 
M_{(x_i,\sigma,i\omega), (x_j,\sigma',i\omega') } =
 \left( i\omega
- t_{x_i,x_j}\right) 
\delta_{\omega,\omega'}  \delta_{\sigma,\sigma'}e^{i\omega'0^+}
+ \frac{\hubbard_{a_i}}{2\beta } \bm{\phi}_{x_i}(i\omega-i\omega')\cdot\bm{\sigma}_{\sigma,\sigma'}e^{i\omega'0^+}
\label{eq:def_M_mat}
\ea
We obtain the following effective action by combining Eq.~\ref{eq:Grassmann_integral}, Eq.~\ref{eq:action_mats} and Eq.~\ref{eq:def_M_mat}
\ba 
S_{eff} = S_\phi -
\text{Tr}[\log(M_{(x_i,\sigma,i\omega), (x_j,\sigma',i\omega') } )]
\ea 
We separate $M$ into two parts 
\ba 
&M_{(x_i,\sigma,i\omega), (x_j,\sigma',i\omega') } = [\tilde{G}^{-1}]_{(x_i,\sigma,i\omega), (x_j,\sigma',i\omega') } + V_{(x_i,\sigma,i\omega), (x_j,\sigma',i\omega') } \nonumber\\ 
&[\tilde{G}^{-1}]_{(x_i,\sigma,i\omega), (x_j,\sigma',i\omega') }  = \left( i\omega
- t_{x_i,x_j}\right) 
\delta_{\omega,\omega'}  e^{i\omega'0^+}\delta_{\sigma,\sigma'}\nonumber\\ 
& V_{(x_i,\sigma,i\omega), (x_j,\sigma',i\omega') } = \delta_{x_i,x_j}\frac{\hubbard_{a_i}}{2\beta } \bm{\phi}_{x_i}(i\omega-i\omega')\cdot\bm{\sigma}_{\sigma,\sigma'}e^{i\omega'0^+}
\label{eq:gree_fun_full}
\ea 
where $\tilde{G}$ is the Green's function of the non-interacting system. 
We can now rewrite the effective action as 
\ba 
\label{eq:eff_action_before_expand}
S_{eff}' = &S_\phi -
\text{Tr}[\log(\tilde{G}^{-1} + V) ] \nonumber\\ 
=&S_\phi -\text{Tr}[\log(\tilde{G}^{-1})] + \sum_{n=1}^{\infty} 
\frac{(-1)^n}{n} \text{Tr}\left[ (\tilde{G}V)^n
\right] 
\ea 
Since $-\text{Tr}[\log(\tilde{G}^{-1})] $ does not depend on the $\bm{\phi}$ fields and is just a constant, 
we drop this term and the final effective action of the $\vec{\phi}$ fields are 
\ba 
S_{eff}= S_\phi+ \sum_{n=1}^{\infty} 
\frac{(-1)^n}{n} \text{Tr}\left[ (\tilde{G}V)^n
\right] 
\label{eq:eff_action_full}
\ea 
We now aim to evaluate the effective action $S_{eff}$ for a system with a narrow band near the Fermi energy.

\section{Interacting narrow bands}
\label{app:set_up_model}
From Eq.~\ref{eq:gree_fun_full}, we find the Green's function can be written as
\ba 
&[\tilde{G}]_{(x_i,\sigma,i\omega), (x_j,\sigma',i\omega') } = G_{x_i,x_i}(i\omega)\delta_{\sigma,\sigma'}\delta_{i\omega,i\omega'}e^{i\omega'0^+} \nonumber\\ 
&G_{x_i,x_j}(i\omega) = \left( i\omega
- t\right)^{-1}_{x_i,x_j}
\ea 
where $t$ is the hopping matrix. 

To evaluate the effective action, we make the following approximations 
\begin{itemize}
    \item 
We separate the Green's function into local $(G_{loc,a_i})$ and non-local part $(G'_{x_i,x_j})$
\ba 
G_{x_i,x_j}(i\omega) 
 = \delta_{x_i,x_j} G_{loc,a_i}(i\omega)  + (1-\delta_{x_i,x_j})G'_{x_i,x_j}(i\omega) 
 \label{eq:green_fun_decomp}
\ea 
As we will discuss in the next section, we treat $G'$ as small parameters and perform an expansion in powers of $G'$.  
We note that, if the non-local part is ignored, the electrons in each unit cell $\RR$ and for each flavor $a$ are decoupled, so their spin orientations can be arbitrary. Only when the non-local contributions are included does the coupling between spin operators of different unit cells or flavors emerge, giving rise to various types of magnetic correlations. 
\item We separate the spin fields $\phi^{\mu}_{\bm{x}_i}(\tau_i)$ into two parts (see Eq.~\ref{eq:spin_field_decomp})
\ba 
\phi^{\mu}_{x} (\tau) =  \phi_{x}^s(\tau) n^\mu_{x}(\tau) 
\ea 
where $\phi_{x}^s(\tau)$ denotes the size of spin moment, and $n^\mu_{x}(\tau)$ denotes the direction of the spin.
% \textcolor{red}{Yes, this is how it was introduced. But we don't need to use $n^\mu_{x}$ time dependent. }
\item In the low-energy limit, due to the strong interaction, we expect the formation of the local moment with
\ba 
\bigg|\bm{\phi}_{x_i}^s(\tau)\bigg| =  \phi_{x_i}^s(\tau) =\phi_{0,a_i} \ne 0
\ea 
Here, we have ignored the dynamical fluctuations ($\tau$-dependency) and spatial dependency ($\RR$-dependency) of $\phi^s_{a_i}(\tau)$ fields. This is because, at low enough temperatures, the size of the local moment is frozen. However, the direction of local moments still fluctuates. We thus drop the $\tau$ and position dependencies of $\phi_x^s(\tau)$ fields but keep them for $n^{\mu}_x(\tau)$ fields. 
This indicates the system stays in a frozen-moment limit where a local moment has developed. However, we also comment that, when other degrees of freedom exist in the system, the local moment could could be Kondo screened. Here, we focus on the magnetic properties of the system and study the formation of magnetic order. 
% \textcolor{red}{We have assumed that there is a frozen moment. At high-T we could have frozen moment in a bad metal, but at low-$T$ such frozen moment is possible only in a selective Mott state, where there is left-over of the local moment, being unscreened by the conduction bands. This is likely the correct ground state at integer fillings with pseudogap. Otherwise Kondo effect should screened the moment out, and make this field zero on average. Note that in heavy fermions this is likely valid only at the critical point, where Fermi liquid does not form, and magnetic long range ordering does not occur.}

% \item We assume there are $n_{flat}$ relatively flat bands near the Fermi energy at the non-interacting limit, whose bandwidths are much smaller than the Hubbard interactions $\hubbard$. 
\end{itemize}

We also discuss the properties of single-particle Green's function. 
We introduce the eigenvalue and eigenbasis of hopping matrix $t_{\kk,ij}$
\ba 
&\sum_b t_{\kk,ab}U_{\kk,bn} = \epsilon_{\kk,n}U_{\kk,an} \nonumber\\ 
&
c_{\kk,j,\sigma} = \sum_n U_{\kk,jn}\gamma_{\kk,n,\sigma}
\ea
where $\gamma$ is the electron operator in the band basis. Then the single-particle Green's function can be calculated via 
\ba 
G_{x_i,x_j}(\tau-\tau') =& -\langle T_\tau c_{x_i,\sigma}(\tau) c_{x_j,\sigma}(\tau')\rangle =  - \frac{1}{N_k}\sum_\kk \langle T_\tau c_{\kk,a_i,\sigma}(\tau) c^\dag_{\kk,a_j,\sigma}(\tau')\rangle e^{i\kk(\RR_j+\bm{r}_{a_j}-\RR_i -\bm{r}_{a_i})} \nonumber\\ 
=&\frac{1}{N} \sum_\kk\sum_n \bigg(-\langle T_\tau \gamma_{\kk,n,\sigma}(\tau) \gamma_{\kk,n,\sigma}^\dag(0)\rangle \bigg) 
U_{\kk, a_in}U_{\kk,a_j n}^* e^{i\kk\cdot(\RR_j+\bm{r}_{a_j}-\RR_i -\bm{r}_{a_i})} .
\ea 
We also note that the Green's function here is the non-interacting Green's function of the system. 
% \textcolor{red}{
% This symbol $U_{\kk,a n}$ can be confused with Coulomb $\hubbard_a$. I am more used to symbol $\psi_{\kk,a n}$, but of course this is a personal preference (from my community).
% I assume the two $\RR+\bm{r}$ stand for vector to the unit cell $\RR$ and vector inside unit cell $\bm{r}_{a}$.
% I would also clarify that this $G$ is actually $G^0$, namely, the Green's function without considering interactions. Because if we have the frozen moment, the interacting Green's function will be very different, namely, the peak that should be at the Fermi level will go into Hubbard bands, and we will be left with a pseudogap. And the single particle self-energy must be almost singular in such a state, namely no quasiparticle is left and quasiparticle $Z=0$ in such a state. 
% }

% where $\bm{r}$ is the position of the sublattice. 
In the Matsubara frequency domain
\ba 
G_{x_j,x_j}(i\omega_n) 
=&\frac{1}{N} \sum_\kk\sum_n \bigg(\frac{1}{i\omega_n-\epsilon_{\kk,n}} \bigg) 
U_{\kk, a_in}U_{\kk,a_j n}^* e^{i\kk\cdot(\RR_j+\bm{r}_{a_j}-\RR_i -\bm{r}_{a_i})} 
\ea 
The local Green's function of sublattice $a$ can also be written as 
\ba 
&G_{loc,a}(i\omega_n) =G_{(\RR,a),(\RR,a)}(i\omega_n)= \int_\epsilon \frac{1}{i\omega_n-\epsilon} \rho_a(\epsilon),\quad \rho_a(\epsilon) =\frac{1}{N} \sum_{\kk,n} |U_{\kk,an}|^2 \delta(\epsilon-\epsilon_{\kk,n})
% = \frac{1}{N} \sum_{\kk,n} |U_{\kk,in}|^2 \delta(\epsilon-\epsilon_{\kk,n})
% \nonumber\\ 
% &G_{\bm{x}_i,\bm{x}_j}(i\omega_n) = 
% -\langle T_\tau c_{\kk,a_i}(\tau) c_{\kk,a_j}(0)\rangle e^
% \frac{1}{N} \sum_{\kk} \frac{U_{\kk,a_in}U_{\kk,a_jn}^*}{i\omega_n -\epsilon_{\kk,n}}e^{i\kk\cdot(\RR_i-\RR_j+\bm{x}_{a_i} -\bm{x}_{a_j})}
\label{eq:def_local_green}
\ea 
where $\rho_a(\epsilon)$ is just the local density of states (DOS). 
We also mention that in multi-orbital systems, where multiple orbitals are located at the same atom, the off-diagonal term $G_{(\RR,a),(\RR,a')}$ with $a\ne a'$ could be sizeable unless specific symmetry enforces it to be zero. 
However, we note that the separation between local and non-local Green's functions is utilized for calculations. 
Treating $G_{(\RR,a),(\RR,a')}$ (for $a \ne a'$) as an off-diagonal component gives rise to an effective spin-spin coupling between spin operators located within the same unit cell $\RR$ but associated with different orbital indices ($a \ne a'$).

We are interested in the case where the non-interacting electrons develop narrow bands near the Fermi energy and produce an enhanced DOS peak. 
We consider the case with $n_{flat}$ narrow bands near the Fermi energy. We separate the dispersion of the $n_{flat}$ narrow bands into two parts
\ba 
\epsilon_{\kk,n} = \epsilon_0 + \delta \epsilon_{\kk,n} ,\quad n=1,...,n_{flat}
\ea 
where 
\ba 
\epsilon_0 = \frac{1}{N}\frac{1}{n_{flat}}\sum_{n,\kk} \epsilon_{\kk,n}
\ea 
represents the average energy of the narrow bands. We consider the case with 
\ba 
\epsilon_0=0
\ea 
such that the narrow bands appear near Fermi energy. Finally, we assume the remote bands are at high energy whose contributions to the Green's function have been ignored.  The local and non-local Green's function can then be written as 
\ba 
&G_{loc,a}(i\omega_n) \approx \frac{1}{N}\sum_{\kk,n=1,...,n_{flat}}\frac{|U_{\kk,an}|^2}{i\omega - \delta \epsilon_{\kk,n}} \nonumber\\
&G'_{x_i,x_j}(i\omega) \approx \frac{1}{N} \sum_{\kk,n=1,...,n_{flat} } \frac{U_{\kk,a_i n}U_{\kk,a_jn}^*}{i\omega -\epsilon_{\kk,n}}e^{i\kk\cdot(\RR_i-\RR_j+\bm{r}_{a_i} -\bm{r}_{a_j})} 
\label{eq:def_non_local_green}
\ea

\section{Effective action}
\label{app:eff_act}
We now derive the effective action by expanding the action in powers of $G'$, which is similar to the gradient expansion. 
The effective action can be written as (from Eq.~\ref{eq:eff_action_full}) 
\ba 
S_{eff}= S_\phi+  \sum_{n=1}^{\infty} 
\frac{(-1)^n}{n} \text{Tr}\left[ (\tilde{G}V)^n
\right] 
\ea 
We let 
\ba 
&\tilde{G} = \tilde{G}_{loc} + \tilde{G}'\nonumber\\ 
&[\tilde{G}_{loc}]_{(x_i,\sigma,i\omega),(x_j,\sigma',i\omega')} = G_{loc,a_i}(i\omega)\delta_{x_i,x_j} \delta_{\sigma,\sigma'}\delta_{\omega,\omega'} e^{i\omega'0^+}\nonumber\\ 
&  [\tilde{G}']_{(x_i,\sigma,i\omega),(x_j,\sigma',i\omega')}  =
G'_{x_i,x_j}(i\omega) e^{i\omega'0^+}\delta_{\sigma,\sigma'}\delta_{\omega,\omega'} \quad \quad \text{ where }x_i\ne x_j
\label{eq:exp_tilde_G}
\ea 
Then by expanding in powers of non-local Green's function $\tilde{G}'$, we observe 
\ba 
S_{eff}&=S_\phi + \sum_{n=1}^{\infty} 
\frac{(-1)^n}{n} \text{Tr}\left[ (\tilde{G}V)^n
\right]  \nonumber\\ 
&\approx   S_0  + S_1 + S_2 \nonumber\\ 
S_0 &= S_\phi +  \sum_{n=1}^{\infty} 
\frac{(-1)^n}{n} \text{Tr}\left[ (\tilde{G}_{loc}V)^n
\right] \nonumber\\ 
S_1&= \sum_{n=1}^{\infty} (-1)^n \text{Tr}
\left[ \tilde{G'}V (\tilde{G}_{loc}V)^{n-1}
\right] \nonumber\\
S_2&= \frac{1}{2}\sum_{n=2}^{\infty} \sum_{k=0}^{n-2}(-1)^n \text{Tr}
\left[ 
 \tilde{G'}V 
 (\tilde{G}_{loc}V)^{k}  \tilde{G'}V 
  (\tilde{G}_{loc}V)^{n-2-k} 
\right]
\label{eq:eff_action_order_exp}
\ea 
where the $S_0,S_1,S_2$ correspond to the zeroth order, first order, and second order contributions (in powers of $\tilde{G}'$), respectively. 
As we discuss in \cref{sec:zeroth_order_term,sec:first_order_term,sec:second_order_term}, the zeroth-order term generates a local on-site contribution, the first-order term vanishes, and the second-order term gives rise to two-body spin-spin interactions. Higher-order terms induce multi-spin interactions beyond the two-body level. For instance, the third-order term leads to three-body interactions of the form $n^\mu_{x_1} n^{\nu}_{x_2} n^{\eta}_{x_3}$.
Since such higher-order terms are generally less relevant in determining the magnetic ordering of the system, we truncate the expansion at second order.

\subsection{$0$-th order term} 
\label{sec:zeroth_order_term}
In this section, we evaluate the 0-th order contribution (Eq.~\ref{eq:eff_action_order_exp})
\ba 
S_0 =& \beta \sum_{x_i}\frac{\hubbard_{a_i}}{4}\phi_{0,a_i}^2  \nonumber\\ 
&+ \sum_{n=1}^\infty 
\frac{(-1)^n}{n}\text{Tr}
\left[ (\tilde{G}_{loc}V)^n\right] 
\ea 
We note that 
\ba 
S_0 &= \beta \sum_{x_i}\frac{\hubbard_{a_i}}{4}\phi_{0,a_i}^2 + \sum_{n=1}^\infty 
\frac{(-1)^n}{n}\text{Tr}
\left[ (\tilde{G}_{loc}V)^n\right] 
= \beta \sum_{x_i}\frac{\hubbard_{a_i}}{4}\phi_{0,a_i}^2 - \text{Tr}[\log( 1 + \tilde{G}_{loc}V)] \nonumber\\ 
&
 = 
  \beta \sum_{x_i}\frac{\hubbard_{a_i}}{4}\phi_{0,a_i}^2 -\text{Tr}[\log(\tilde{G}_{loc})]- \text{Tr}[\log( \tilde{G}_{loc}^{-1} + V)]
  \label{eq:s0_define}
\ea 
We can again use the Gaussian integrals
\ba 
\int D[\eta,\eta^\dag] e^{\sum_{ij}\eta_i^\dag  M_{ij} \eta_j } = e^{\text{Tr}[\log(M)]}
\ea 
We then find
\ba 
&e^{ \text{Tr}[\log( \tilde{G}_{loc}^{-1} + V)]} \nonumber\\ 
=&\prod_{x_i}\int D[\eta^\dag_{x_i}, \eta_{x_i}]
e^{+
\int_{\tau,\tau'}\sum_\sigma \eta_{x_i,\sigma}^\dag(\tau) G^{-1}_{loc,a_i}(\tau-\tau') \eta_{x_i,\sigma}(\tau') + \int_\tau \sum_{\sigma,\sigma'}\frac{\hubbard_{a_i}\bm{\phi}_{x_i}(\tau) \cdot \bm{\sigma}_{\sigma,\sigma'}}{2}\delta(\tau-\tau')\eta_{x_i,\sigma}^\dag(\tau) \eta_{x_i,\sigma'}(\tau') 
}
\label{eq:local_action_fermi_rep}
\ea 
We note that the local DOS of the non-interacting system develops a peak at the Fermi energy due to the existence of the narrow bands. Approximately, we have 
\ba  
\rho_a(\epsilon) \approx \frac{1}{N}\sum_{\kk,n=1,...,n_{flat}}|U_{\kk,an}|^2 \delta(\epsilon - \delta_{\epsilon_{\kk,n}} )\approx A_a\delta(\epsilon) 
\label{eq:flat_band_norm_cond_1}
\ea 
where the prefactor is defined as 
\ba 
A_a = \frac{1}{N}\sum_\kk \sum_{n=1,...,n_{flat}}|U_{\kk,an}|^2
\label{eq:def_para_A}
\ea 
Then the local Green's function can be approximately written as 
\ba 
G_{loc,a}(i\omega)\approx \frac{A_a}{i\omega}
\label{eq:local_green_fun}
 \ea

Using the~\cref{eq:exp_tilde_G,eq:local_green_fun,eq:local_action_fermi_rep}, we find 
\ba 
&e^{ \text{Tr}[\log( \tilde{G}_{loc}^{-1} + V)]}  \nonumber\\ 
=& 
\prod_{x_i}\int D[\eta_{x_i,\sigma}^\dag(\tau), \eta_{x_i,\sigma}(\tau)]\exp\left\{ - 
\int_{\tau,\tau'} \eta_{x_i,\sigma}^\dag(\tau)  A_{a_i}^{-1}\left[\delta_{\sigma,\sigma'}\partial_\tau -
g_{a_i}\bm{n}_{x_i}(\tau) \cdot \bm{\sigma}_{\sigma,\sigma'}\right] 
\eta_{x_i,\sigma'}(\tau) 
\right\} 
\ea 
where we have introduced 
\ba 
g_{a_i} = \phi_{0,a_i}\hubbard_{a_i}A_{a_i}/2.
\ea 
% \ba 
% e^{- \text{Tr}[\log( \tilde{G}_{loc}^{-1} + V)]}  = \int D[\eta,\eta^\dag]
% \exp\bigg\{ 
% - \int d\tau \sum_{\sigma',\sigma'',\mu}c_{\RR+\rr_a,\sigma'}^\dag(\tau)\bigg( A_a^{-1}
% \delta_{\sigma',\sigma''}(\partial_\tau +\epsilon_0) -
% g_an_{\RR+\rr_a}^\mu(\tau)\sigma^\mu_{\sigma',\sigma''} \bigg)c_{\RR+\rr_a,\sigma''}(\tau)
% \bigg\} 
% \ea 
We now evaluate this term explicitly. We first introduce the following parametrization of the $n$ fields
\ba 
n_{x_i}^\mu(\tau) = \begin{bmatrix}
    \sin(\theta_{x_i}(\tau)) \cos(\chi_{x_i}(\tau)) & \sin(\theta_{x_i}(\tau)) \sin(\chi_{x_i}(\tau)) & \cos(\theta_{x_i}(\tau))
\end{bmatrix}
\ea 
We introduce the fluctuation frame and define new fermionic fields $\psi$, which are standard Grassmann variables
\ba 
&R_{x_i}(\tau) = 
\begin{bmatrix}
    -e^{-i\chi_{x_i}(\tau)} \sin(\frac{\theta_{x_i}(\tau)}{2}) & e^{-i\chi_{x_i}(\tau)}\cos(\frac{\theta_{x_i}(\tau)}{2}) \\
    \cos(\frac{\theta_{x_i}(\tau)}{2} )& \sin(\frac{\theta_{x_i}(\tau)}{2})
\end{bmatrix} \nonumber\\ 
&\eta_{{x_i},\sigma}(\tau) = \sqrt{A_{a_i}}\sum_{\sigma'}[R_{x_i}(\tau)]_{\sigma,\sigma'}\psi_{{x_i},\sigma'}(\tau)
\ea 
The action now behaves as 
\ba 
&e^{ \text{Tr}[\log( \tilde{G}_{loc}^{-1} + V)]} \nonumber\\ 
=&\prod_{x_i}\int D[\psi, \psi^\dag]
\exp\Bigg\{ 
- \int d\tau \sum_{\sigma',\sigma''}\psi_{{x_i},\sigma'}^\dag(\tau)\bigg\{
\delta_{\sigma',\sigma''}\partial_\tau 
+\bigg[
R_{x_i}^\dag(\tau) \partial_\tau R_{x_i}(\tau)
\bigg]_{\sigma',\sigma''}
+\sigma' 
g_{a_i}\delta_{\sigma',\sigma''}
\bigg\}  \psi_{x_i,\sigma''}(\tau) \Bigg\}
\ea 
We can evaluate the integral in the power series of $R_{x_i}^\dag(\tau) \partial_\tau R_{x_i}(\tau)$ (which is equivalent to a gradient expansion in powers of $\partial_\tau$). We only keep the zeroth order and first order term 
\ba 
&e^{ \text{Tr}[\log( \tilde{G}_{loc}^{-1} + V)]} \nonumber\\ 
=&\prod_{x_i}\int D[\psi,\psi^\dag] 
\exp\Bigg\{ 
- \int d\tau \sum_{\sigma',\sigma''}\psi_{x_i,\sigma'}^\dag(\tau)\bigg\{ 
\delta_{\sigma',\sigma''}\partial_\tau 
+\bigg[
R_{x_i}^\dag(\tau) \partial_\tau R_{x_i}(\tau)
\bigg]_{\sigma',\sigma''}
+\sigma' 
g_{a_i}\delta_{\sigma',\sigma''}
\bigg\}  \psi_{{x_i},\sigma''}(\tau) \Bigg\} \nonumber\\ 
\approx  &\int D[\psi,\psi^\dag] 
\bigg\{ 1 - \int_\tau d\tau \sum_{\sigma',\sigma''}\bigg[
R_{x_i}^\dag(\tau) \partial_\tau R_{x_i}(\tau)
\bigg]_{\sigma',\sigma''}\psi_{{x_i},\sigma'}^\dag(\tau)\psi_{{x_i},\sigma''}(\tau) 
\bigg\} 
\nonumber\\ 
&
\exp\Bigg\{ 
- \int d\tau \sum_{\sigma}\psi_{{x_i},\sigma}^\dag(\tau)\bigg\{ 
\partial_\tau 
+\sigma 
g_{a_i}
\bigg\}  \psi_{{x_i},\sigma}(\tau) \Bigg\}\\ 
\approx &
Z_{0,loc} - Z_{0,loc}\int_\tau \sum_{\sigma',\sigma''}\bigg[
R_{{x_i}}^\dag(\tau) \partial_\tau R_{{x_i}}(\tau)
\bigg]_{\sigma',\sigma''}\bigg\langle \psi_{{x_i},\sigma'}^\dag(\tau)\psi_{{x_i},\sigma''}(\tau) 
\bigg\rangle_{0,loc} 
\label{eq:seff_loc}
\ea

where the partition function and the expectation value (for a given operator $O$) are defined as 
\ba 
Z_{0,loc} =& \prod_{x_i}\int D[\psi,\psi^\dag]
\exp\Bigg\{ 
- \int d\tau \sum_{\sigma',\sigma''}\psi_{{x_i},\sigma'}^\dag(\tau)\bigg\{ 
\delta_{\sigma',\sigma''}\partial_\tau 
+\sigma' 
g_{a_i}\delta_{\sigma',\sigma''}
\bigg\}  \psi_{x_i,\sigma''}(\tau) \Bigg\} \nonumber\\ 
\langle O \rangle_{0,loc}=&\frac{1}{Z_{0,loc}}\prod_{x_i}\int D[\psi,\psi^\dag]O
\exp\Bigg\{ 
- \int d\tau \sum_{\sigma',\sigma''}\psi_{{x_i},\sigma'}^\dag(\tau)\bigg\{ 
\delta_{\sigma',\sigma''}\partial_\tau 
+\sigma' 
g_{a_i}\delta_{\sigma',\sigma''}
\bigg\}  \psi_{x_i,\sigma''}(\tau) \Bigg\}
\label{eq:local_theory_exp}
\ea

Now $Z_{0,loc,\RR}$ is just the partition function of a non-interacting system with energy 
\ba 
 E_{loc,a_i,\pm} = \pm  g_{a_i} 
\ea 
The partition function of this non-interacting system is 
\ba 
Z_{0,loc} = \prod_{x_i}\bigg[ \bigg( 1+e^{-\beta E_{loc,a_i,+}}
\bigg) \bigg( 1+e^{\beta E_{loc,a_i,-}}
\bigg) \bigg] 
\label{eq:z_loc}
\ea 
Moreover, the expectation values are 
\ba 
\langle \psi_{x_i,\sigma'}^\dag(\tau) \psi_{x_i,\sigma''}(\tau) \rangle_{0,loc}  =\delta_{\sigma',\sigma''} n_F(E_{loc,a_i,\sigma'})
\ea 
where the Fermi-Dirac function is $n_F(x) =1/(1+e^{\beta \epsilon})$. At the low-temperature limit, we find 
\ba 
\langle \psi_{x_i,\sigma'}^\dag(\tau) \psi_{x_i,\sigma''}(\tau) \rangle_{0,loc} = \delta_{\sigma',\sigma''}\frac{1-\sigma'}{2}
\ea 

Then the first-order contribution in Eq.~\ref{eq:local_theory_exp} gives
\ba 
 &- Z_{0,loc}\int_\tau \sum_{\sigma',\sigma''}\sum_{x_i}\bigg[
R_{{x_i}}^\dag(\tau) \partial_\tau R_{{x_i}}(\tau)
\bigg]_{\sigma',\sigma''}\bigg\langle \psi_{{x_i},\sigma'}^\dag(\tau)\psi_{{x_i},\sigma''}(\tau) 
\bigg\rangle_{0,loc} \nonumber\\ 
=& -iZ_{0,loc} \int_\tau \sum_{x_i} \frac{-\cos(\theta_{x_i}(\tau))-1}{2}\partial_\tau \chi_{x_i}(\tau) \nonumber\\ 
=&-iZ_{0,loc}\sum_{x_i}A[\bm{n}_{x_i}]
\label{eq:berry_phase_contrib}
\ea 
where we define
\ba 
A[\bm{n}_{x_i}]= -\int_\tau \frac{\cos(\theta_{x_i}(\tau))+1}{2}\partial_\tau \phi_{x_i}(\tau) 
\ea 
which is just the conventional Berry phase term of a spin system.  
Then the effective action now reads (from Eq.~\ref{eq:seff_loc}, Eq.~\ref{eq:z_loc}, and Eq.~\ref{eq:berry_phase_contrib})  
\ba 
&e^{ \text{Tr}[\log( \tilde{G}_{loc}^{-1} + V)]} \approx e^{-\left[   -\log 
\bigg(Z_{0,loc}(1- iA[\bm{n}_{x_i}])\bigg) \right] } \nonumber\\ 
\Rightarrow &  -\text{Tr}[\log( \tilde{G}_{loc}^{-1} + V)]
\approx -\log Z_{0,loc}  +iA[\bm{n}_{x_i}]
\label{eq:loc_act}
\ea 
Combining Eq.~\ref{eq:s0_define} and Eq.~\ref{eq:loc_act}, we find
\ba 
S_0 =\sum_{x_i}\left[ \beta \sum_{x_i}\frac{\hubbard_{a_i}}{4}\phi_{0,a_i}^2 -\log Z_{0,loc}  +iA[\bm{n}_{x_i}] \right] -\text{Tr}[\log\tilde{G}_{loc}] 
\label{eq:s0_eval}
\ea  
where $\text{Tr}[\log\tilde{G}_{loc}] $ is just a constant. 

We note that $S_0$ represents the zeroth-order (in $\tilde{G}'$) term in the action expansion (Eq.~\eqref{eq:eff_action_order_exp}). It consists of two contributions: the action of the $\phi_{0,a}$ field, denoted by $S_{\phi_0}$, and the Berry phase term, $iA[\bm{n}_{x_i}]$. 
More explicitly, we have 
\ba 
S_{\phi_{0} }&= -\log(Z_{0,loc}) +S_\phi = \sum_{x_i}\bigg\{ \beta \frac{\hubbard_{a_i}}{4}\phi_{0,a_i}^2  -
\log 
\bigg[\bigg(1+e^{-\beta (\hubbard_{a_i}A_{a_i}\phi_{0,{a_i}}/2)}\bigg) 
\bigg(1+e^{\beta ( \hubbard_{a_i}A_{a_i}\phi_{0,{a_i}}/2)}\bigg) 
\bigg] \bigg\} \nonumber\\
&= \sum_{x_i}\bigg\{ \beta \frac{\hubbard_{a_i}}{4}\phi_{0,a_i}^2  -
2\log 
\bigg[2\cosh\left(\frac{1}{4}\beta \hubbard_{a_i}A_{a_i}\phi_{0,{a_i}}\right) 
\bigg] \bigg\} 
\ea 

The value of $\phi_{0,{a_i}}$ and be determined by the saddle-point equation
\ba
\frac{\delta S_{\phi_{0}}}{\delta \phi_{0,{a_i}} } =0 
\Rightarrow \beta \frac{1}{2} \hubbard_{a_i} \phi_{0,{a_i}} - 
\frac{1}{2} \beta \hubbard_{a_i}A_{a_i} \tanh(\frac{\beta \hubbard_{a_i}A_{a_i}\phi_{0,{a_i}}}{4}
)=0
\label{eq:sad_eq_phi0}
\ea

The saddle point equation has a non-zero solution only below $T_c$. We can determine $T_c$ by calculating the mass term of $\phi_{0,{a_i}}$
\ba 
m_{a_i} = \frac{\delta^2 S_{\phi_{0}}}{\delta \phi_{0,{a_i}}^2 }\bigg|_{\phi_{0,{a_i}}=0} 
= \frac{\beta \hubbard_{a_i}}{2} -
\frac{\beta^2 A_{a_i}^2\hubbard_{a_i}^2}{8} 
\ea 
% \textcolor{red}{
% Or we can just find solution for small $\phi_{0,{a_i}}$, at which $\tanh x\approx x$ and we get the same $T_c$.
% At low $T$ we have $\tanh x=1$, hence $\phi_{0,{a_i}}=A_{a_i}$ as found below.
% }

The $T_{c,{a_i}}$ corresponds to the temperature below which $m_{a_i}<0$. We find
\ba 
T_{c,{a_i}} = \frac{ A_{a_i}^2}{4}\hubbard_{a_i}
\ea 
$T_{c,a_i}$ is then the temperature below which the sublattice $a_i$ starts to develop local moment ($\phi_{0,a}\ne 0$).  

We are mostly interested in the low-temperature limit. At the low-temperature limit, we can directly solve Eq.~\ref{eq:sad_eq_phi0} which gives 
\ba 
& \frac{1}{2} \hubbard_{a_i} \phi_{0,{a_i}} - 
\frac{1}{2} \hubbard_{a_i}A_{a_i} \tanh(\frac{\beta \hubbard_{a_i}A_{a_i}\phi_{0,{a_i}}}{4}
)=0 \nonumber\\ 
\Rightarrow & \phi_{0,a_i} = A_{a_i }\tanh(\frac{\beta \hubbard_{a_i}A_{a_i}\phi_{0,{a_i}}}{4}) 
\ea 
At low-temperature limit with $\beta \rightarrow \infty$, we have $\tanh(\frac{\beta \hubbard_{a_i}A_{a_i}\phi_{0,{a_i}}}{4}) \rightarrow 1 $, and then 
\ba 
\phi_{0,{a_i}} =A_{a_i}
\label{eq:val_phi0}
\ea 
In other words, the size of the local moment is $A_{a_i}$, which is also dimensionless.

The contribution to the effective action of the $\bm{n}_x$ fields is just a simple Berry phase term which takes the form of 
\ba 
S_{B} = i \sum_{x}A[n_{x}] 
\label{eq:berry_phase}
\ea 

Finally, we also comment on the phase transition suggested by the current calculation, which separates a high-temperature regime without local moment formation and a low-temperature phase with the development of local moment. Such a sharp transition could be an artifact of the current expansion. However, $T_{c,a_i}$ can still be understood as the energy scale where the local moment behaviors start to appear.

\subsection{First-order term}
\label{sec:first_order_term}
We now discuss the first-order term $S_1$ in Eq.~\ref{eq:eff_action_order_exp}. 
\ba 
S_1=& \sum_{n=1}^{\infty} (-1)^n \text{Tr}
\left[ \tilde{G'}V (\tilde{G}_{loc}V)^{n-1}
\right] \nonumber\\ 
=& 
\sum_{n=1}^{\infty} \sum_{i\omega_1,...,i\omega_{n}=i\omega_1}\sum_{\sigma_1,...,\sigma_n=\sigma_1}(-1)^n \nonumber\\ 
&\text{Tr}
\left[ [\tilde{G'}]_{(x,\sigma_1,i\omega_1), (x,\sigma_1,i\omega_1)} \prod_{m=1}^{n-1}[V]_{(x,\sigma_m,i\omega_m),(x,\sigma_{m+1},i\omega_{m+1})}[\tilde{G}_{loc}]_{(x,\sigma_{m+1},i\omega_{m+1}),(x,\sigma_{m+1},i\omega_{m+1})}\right] 
\ea 
where we have use the fact that $V $ and $\tilde{G}_{loc}$ are diagonal with respect to the position index $x$ (see Eq.~\ref{eq:gree_fun_full} and Eq.~\ref{eq:exp_tilde_G}). In addition, since $\tilde{G'}$ represents the non-local component of the Green's function, $ [\tilde{G'}]_{(x,\sigma_1,i\omega_1), (x,\sigma_1,i\omega_1)}=0$. Therefore, we conclude
\ba 
S_1=0 
\ea

\subsection{Second-order term}
\label{sec:second_order_term}
We now discuss the second-order term $S_2$ in Eq.~\ref{eq:eff_action_order_exp}. By combining Eq.~\ref{eq:gree_fun_full}, Eq.~\ref{eq:exp_tilde_G} and Eq.~\ref{eq:eff_action_order_exp}, we observe 
\ba 
S_2 =& \frac{1}{2}\sum_{n=2}^{\infty}\sum_{k=0}^{n-2}(-1)^n 
\sum_{x_i,x_j}\sum_{i\omega,i\omega',i\tilde{\omega},i\tilde{\omega'},i\omega'_1,..,i\omega'_{k+1}, i\omega_1,...,i\omega_{n-1-k}} 
\sum_{\sigma,\sigma', \tilde{\sigma},\tilde{\sigma'},\sigma_1,...,i\sigma_{k+1},\sigma'_1,...,i\sigma'_{n-1-k} }\nonumber\\
&\delta_{i\omega', i\omega_{1}'}\delta_{i\omega'_{k+1},i\tilde{\omega}}\delta_{i\tilde{\omega}',i\omega_1}
\delta_{i\omega_{n-1-k},i\omega} 
\delta_{\sigma',\sigma_1'}\delta_{\sigma_{k+1}',\tilde{\sigma}}
\delta_{\tilde{\sigma}',\sigma_{1}}\delta_{\sigma_{n-1-k},\sigma}e^{i\omega0^+}
\nonumber\\ 
&
G'_{x_i,x_j}(i\omega)  \frac{\hubbard_{a_j}\bm{\phi}_{x_j}(i\omega-i\omega')\cdot \bm{\sigma}_{\sigma,\sigma'}}{2\beta}
\left( \prod_{m=1}^{k}
[G_{loc,a_j}(i\omega'_m)]\frac{\hubbard_{a_j}
\bm{\phi}_{x_j}(i\omega'_m-i\omega'_{m+1})\cdot \bm{\sigma}_{\sigma'_m,\sigma'_{m+1}}
}{2\beta}
\right)  \nonumber\\ 
&G'_{x_j,x_i}(i\tilde{\omega})\frac{\hubbard_{a_i}\bm{\phi}_{x_i}(i\tilde{\omega}-i\tilde{\omega'})\cdot \bm{\sigma}_{\tilde{\sigma},\tilde{\sigma'}}}{2\beta}
\left( 
\prod_{s=1}^{n-2-k}
[G_{loc,a_i}(i\omega_s)]
% \frac{A_{a_i}}{i\omega_s}
\frac{\hubbard_{a_i}
\bm{\phi}_{x_i}(i\omega_s-i\omega_{s+1})\cdot \bm{\sigma}_{\sigma_s,\sigma_{s+1}}
}{2\beta}
\right) \nonumber\\ 
=& \frac{1}{2}
\sum_{n=2}^{\infty}\sum_{k=0}^{n-2}(-1)^n 
\sum_{x_i,x_j}\sum_{i\omega,i\Omega,i\Omega',i\Omega_1',...,i\Omega_k',i\Omega_1,..,i\Omega_{n-2-k} }
\sum_{\sigma,\sigma', \tilde{\sigma},\tilde{\sigma'},\sigma_1,...,i\sigma_{k+1},\sigma'_1,...,i\sigma'_{n-1-k} }\nonumber\\
&\delta_{i\Omega+i\Omega'+\sum_{m=1}^ki\Omega'_m +\sum_{s=1}^{n-2-k}i\Omega_s, 0} 
\delta_{\sigma',\sigma_1'}\delta_{\sigma_{k+1}',\tilde{\sigma}}
\delta_{\tilde{\sigma}',\sigma_{1}}\delta_{\sigma_{n-1-k},\sigma} e^{i\omega0^+}
\nonumber\\ 
&
G'_{x_i,x_j}(i\omega)  \frac{\hubbard_{a_j}\bm{\phi}_{x_j}(i\Omega)\cdot \bm{\sigma}_{\sigma,\sigma'}}{2\beta}
\left( \prod_{m=1}^{k}
[G_{loc,a_j}(i\omega-i\Omega-\sum_{t=1}^{m-1}i\Omega'_t)]
\frac{\hubbard_{a_j}
\bm{\phi}_{x_j}(i\Omega'_{m})\cdot \bm{\sigma}_{\sigma'_m,\sigma'_{m+1}}
}{2\beta}
\right)  \nonumber\\ 
&G'_{x_j,x_i}(i\omega-i\Omega-\sum_{m=1}^ki\Omega'_m)\frac{\hubbard_{a_i}\bm{\phi}_{x_i}(i\Omega')\cdot \bm{\sigma}_{\tilde{\sigma},\tilde{\sigma'}}}{2\beta}
\left( 
\prod_{s=1}^{n-2-k}
[G_{loc,a_i}(i\omega-i\Omega-\sum_{m=1}^ki\Omega'_m-i\Omega'-\sum_{t=1}^{s-1}i\Omega_t )]\frac{\hubbard_{a_i}
\bm{\phi}_{x_i}(i\Omega_s)\cdot \bm{\sigma}_{\sigma_s,\sigma_{s+1}}
}{2\beta}
\right)  \nonumber\\ 
=&  \frac{1}{2} \sum_{n=2}^\infty \sum_{k=0}^{n-2} (-1)^n 
\sum_{x_i,x_j} F_{x_i,x-j}(i\Omega,i\Omega_1',..,i\Omega_k',i\Omega',i\Omega_1,...,i\Omega_{n-2-k}) \delta_{i\Omega+i\Omega'+\sum_{m=1}^ki\Omega'_m +\sum_{s=1}^{n-2-k}i\Omega_s, 0} \nonumber\\ 
& \text{Tr}\left[ \frac{\hubbard_{a_j}
\bm{\phi}_{x_j}(i\Omega)\cdot \bm{\sigma}}
{2\beta}
\left( 
\prod_{m=1}^{k}\frac{\hubbard_{a_j}
\bm{\phi}_{x_j}(i\Omega'_{m})\cdot \bm{\sigma}
}{2\beta}\right) 
\frac{\hubbard_{a_i}
\bm{\phi}_{x_i}(i\Omega)\cdot \bm{\sigma}
}{2\beta}\left(
\prod_{s=1}^{n-2-k}\frac{\hubbard_{a_i}
\bm{\phi}_{x_i}(i\Omega_s)\cdot \bm{\sigma}
}{2\beta}\right) 
\right]  
\ea 
where the interaction vertex is defined as
\ba 
&F_{x_i,x_j}(i\Omega,i\Omega_1',..,i\Omega_k',i\Omega',i\Omega_1,...,i\Omega_{n-2-k}) \nonumber\\
= &\sum_{i\omega}
G'_{x_i,x_j}(i\omega)  
\left( \prod_{m=1}^{k}
[G_{loc,a_j}(i\omega-i\Omega-\sum_{t=1}^{m-1}i\Omega'_t)]
\right) G'_{x_j,x_i}(i\omega-i\Omega-\sum_{m=1}^ki\Omega'_m) \nonumber\\ 
&
\left( 
\prod_{s=1}^{n-2-k}[G_{loc,a_i}(i\omega-i\Omega-\sum_{m=1}^ki\Omega'_m-i\Omega'-\sum_{t=1}^{s-1}i\Omega_t )]
\right)  e^{i\omega0^+}
\ea 
We take the low-frequency limit of the interaction vertex by evaluating it at zero frequency. In other words, we neglect its frequency dependence and approximate the vertex by its value at zero frequency
\ba 
&F_{x_i,x_j}(i\Omega,i\Omega_1',..,i\Omega_k',i\Omega',i\Omega_1,...,i\Omega_{n-2-k})  
\approx   F_{x_i,x_j}(0,0,...,0,0,0,...,0) 
\ea 
We then obtain the following contributions 
\ba 
S_2^{(0)} \approx &\frac{1}{2}
 \sum_{n=2}^\infty \sum_{k=0}^{n-2} (-1)^n 
\sum_{x_i,x_j}
\sum_{i\omega}\left[G_{loc,a_j}(i\omega)\right]^k \left[G_{loc,a_i}(i\omega)\right]^{n-2-k} G'_{x_i,x_j}(i\omega)G'_{x_j,x_j}(i\omega) e^{i\omega0^+}
\delta_{i\Omega+i\Omega'+\sum_{m=1}^ki\Omega'_m +\sum_{s=1}^{n-2-k}i\Omega_s, 0} \nonumber\\ 
& \text{Tr}\left[ \frac{\hubbard_{a_j}
\bm{\phi}_{x_j}(i\Omega)\cdot \bm{\sigma}}
{2\beta}
\left( 
\prod_{m=1}^{k}\frac{\hubbard_{a_j}
\bm{\phi}_{x_j}(i\Omega'_{m})\cdot \bm{\sigma}
}{2\beta}\right) 
\frac{\hubbard_{a_i}
\bm{\phi}_{x_i}(i\Omega)\cdot \bm{\sigma}
}{2\beta}\left(
\prod_{s=1}^{n-2-k}\frac{\hubbard_{a_i}
\bm{\phi}_{x_i}(i\Omega_s)\cdot \bm{\sigma}
}{2\beta}\right) 
\right]  
\ea 

We then transform the bosonic field $\bm{\phi}_{x_j}(i\Omega)$ to the imaginary time domain which gives 
\ba 
S_2^{(0)} \approx &
  \sum_{n=2}^\infty \sum_{k=0}^{n-2} (-1)^n 
\sum_{x_i,x_j}
\sum_{i\omega}\left(\frac{\hubbard_{a_j}}{2}[G_{loc,a_j}(i\omega)]\right)^k \left(\frac{\hubbard_{a_i} \phi_{0,a_i}}{2}[G_{loc,a_i}(i\omega)]\right)^{n-2-k} \frac{\hubbard_{a_j}\hubbard_{a_i}}{4}G'_{x_i,x_j}(i\omega)G'_{x_j,x_j}(i\omega) e^{i\omega0^+}\nonumber\\ 
& \frac{1}{2\beta} \int_\tau \text{Tr}\left[ 
\bm{\phi}_{x_j}(\tau)\cdot \bm{\sigma}
\left( 
\prod_{m=1}^{k}
\bm{\phi}_{x_j}(\tau)\cdot \bm{\sigma}\right) 
\bm{\phi}_{x_i}(\tau)\cdot \bm{\sigma}\left(
\prod_{s=1}^{n-2-k}
\bm{\phi}_{x_i}(\tau)\cdot \bm{\sigma}
\right) 
\right] 
\ea 
We then use 
\ba 
&\bm{\phi}_{x_j}(\tau) = \phi_{0,a_j}\bm{n}_{x_j}(\tau) \nonumber\\ 
&
\sum_{\mu_1,\mu_2}n_{x_i}^{\mu_1}(\tau) n_{x_i}^{\mu_2}(\tau) \bigg(\sigma^{\mu_1}\cdot\sigma^{\mu_2}\bigg) = |\bm{n}_{x_i}|^2 \mathbb{I} = \mathbb{I} \nonumber\\ 
&\text{Tr}\bigg[ (\bm{n}_{x_i}(\tau)\cdot \bm{\sigma})(\bm{n}_{x_j}(\tau)\cdot \bm{\sigma})  \bigg] = 2 \bm{n}_{x_i}(\tau)\cdot \bm{n}_{x_j}(\tau)
\ea 
and find 
\ba 
S_2^{(0)} \approx &
 - \sum_{n=2}^\infty \sum_{k=0}^{n-2} (-1)^n 
\sum_{x_i,x_j}
\sum_{i\omega}\left(\frac{\hubbard_{a_j} \phi_{0,a_j}}{2}[G_{loc,a_j}(i\omega)]\right)^k \left(\frac{\hubbard_{a_i} \phi_{0,a_i}}{2}[G_{loc,a_i}(i\omega)]\right)^{n-2-k} \frac{\hubbard_{a_j}\hubbard_{a_i}\phi_{0,a_j}\phi_{0,a_i}}{8} \nonumber\\
&G'_{x_i,x_j}(i\omega)G'_{x_j,x_j}(i\omega) e^{i\omega0^+} \frac{1}{\beta} \int_\tau \text{Tr}\left[ 
\left(
\bm{n}_{x_j}(\tau)\cdot \bm{\sigma}\right)^{k+1} \left(
\bm{n}_{x_i}(\tau)\cdot \bm{\sigma}
\right)^{n-1-k} 
\right]  \nonumber\\ 
=& 
 - \sum_{n=2}^\infty \sum_{k=0}^{n-2} (-1)^n 
\sum_{x_i,x_j}
\sum_{i\omega}\left(\frac{\hubbard_{a_j} \phi_{0,a_j}}{2}[G_{loc,a_j}(i\omega)]\right)^k \left(\frac{\hubbard_{a_i} \phi_{0,a_i}}{2}[G_{loc,a_i}(i\omega)]\right)^{n-2-k} \frac{\hubbard_{a_j}\hubbard_{a_i}\phi_{0,a_j}\phi_{0,a_i}}{4} \nonumber\\
&G'_{x_i,x_j}(i\omega)G'_{x_j,x_j}(i\omega)e^{i\omega0^+} \frac{1}{\beta} \int_\tau \left[ \frac{(-1)^{k+1}+1}{2}
\frac{(-1)^{n-1-k}+1}{2} + 
\frac{(-1)^{k}+1}{2}
\frac{(-1)^{n-k}+1}{2}\bm{n}_{x_i}(\tau)\cdot \bm{n}_{x_j}(\tau)\right] 
\ea 
Since we aim to derive effective spin-spin interaction terms, we can drop the term that does not depend on $\bm{n}_{x_i}(\tau)$ fields. The remaining terms are
\ba 
S_2^{(0)\prime }= &
- \sum_{n=2}^\infty \sum_{k=0}^{n-2} (-1)^n 
\sum_{x_i,x_j}
\sum_{i\omega}\left(\frac{\hubbard_{a_j} \phi_{0,a_j}}{2}[G_{loc,a_j}(i\omega)]\right)^k \left(\frac{\hubbard_{a_i} \phi_{0,a_i}}{2}[G_{loc,a_i}(i\omega)]\right)^{n-2-k} \frac{\hubbard_{a_j}\hubbard_{a_i}\phi_{0,a_j}\phi_{0,a_i}}{4} \nonumber\\
&\frac{(-1)^{k}+1}{2}
\frac{(-1)^{n-k}+1}{2}G'_{x_i,x_j}(i\omega)G'_{x_j,x_j}(i\omega) e^{i\omega0^+}\frac{1}{\beta} \int_\tau \left[ \bm{n}_{x_i}(\tau)\cdot \bm{n}_{x_j}(\tau)\right] 
\ea 
We can let $ m= n-2-k$ and replace the summation over $n,k$ by the summation over $k,m$
\ba 
S_2^{(0)\prime}= &-\sum_{k=0}^{\infty} \sum_{m=0}^{\infty} 
 (-1)^{k+m}
\sum_{x_i,x_j}
\sum_{i\omega}\left(\frac{\hubbard_{a_j} \phi_{0,a_j}}{2}[G_{loc,a_j}(i\omega)]\right)^k \left(\frac{\hubbard_{a_i} \phi_{0,a_i}}{2}[G_{loc,a_i}(i\omega)]\right)^{m} \frac{\hubbard_{a_j}\hubbard_{a_i}\phi_{0,a_j}\phi_{0,a_i}}{4}\frac{(-1)^{k}+1}{2}
\frac{(-1)^{m}+1}{2}\nonumber\\ 
& G'_{x_i,x_j}(i\omega)G'_{x_j,x_j}(i\omega)e^{i\omega0^+} \frac{1}{\beta} \int_\tau \left[ \bm{n}_{x_i}(\tau)\cdot \bm{n}_{x_j}(\tau)\right] \nonumber\\ 
=& -
\sum_{x_i,x_j}\frac{1}{\beta} \sum_{i\omega} \frac{1}{1-\left(\frac{A_{a_i}\hubbard_{a_i}\phi_{0,a_i}}{2i\omega}\right)^2} \frac{1}{1-\left(\frac{A_{a_j}\hubbard_{a_j}\phi_{0,a_j}}{2i\omega}\right)^2}\frac{\hubbard_{a_j}\hubbard_{a_i}\phi_{0,a_j}\phi_{0,a_i}}{4} G'_{x_i,x_j}(i\omega)G'_{x_j,x_j}(i\omega)e^{i\omega0^+} \int_\tau \left[ \bm{n}_{x_i}(\tau)\cdot \bm{n}_{x_j}\right] \nonumber\\ 
=& \int_\tau \sum_{x_i,x_j } J_{x_i,x_j}\bm{n}_{x_i}(\tau)\cdot \bm{n}_{x_j}(\tau) 
\label{eq:seff_spin_spin}
\ea 
where the effective spin-spin interactions are defined as 
\ba 
\label{eq:J_xixj_from_Green}
J_{x_i,x_j} = \frac{1}{\beta} \sum_{i\omega} \frac{1}{1-\left(\frac{\hubbard_{a_i} A_{0,a_i}}{2}[G_{loc,a_i}(i\omega)]\right)^2} \frac{1}{1-\left(\frac{\hubbard_{a_j} A_{0,a_j}}{2}[G_{loc,a_j}(i\omega)]\right)^2}\frac{\hubbard_{a_j}\hubbard_{a_i}A_{a_j}A_{a_i}}{4} G'_{x_i,x_j}(i\omega)G'_{x_j,x_j}(i\omega) e^{i\omega0^+}
\ea 
where we have also replace $\phi_{0,a_i}$ by its saddle-point value $A_{a_i}$ (Eq.~\ref{eq:val_phi0}).

The spin–spin interaction can be further simplified and expressed explicitly in terms of the wavefunctions and energy dispersion of the non-interacting bands.
From \cref{eq:J_xixj_from_Green}, we proceed under the following assumption: 
\begin{itemize}
    \item We assume all the electronic orbitals are equivalent. This indicates 
    \ba 
    \hubbard_{a_i}=\hubbard 
    \ea 
    \item We expand the expression (\cref{eq:J_xixj_from_Green}) in powers of $\delta \epsilon_{\kk,n}$. Such an expansion can first be performed at the Green's function level, which gives 
    \ba 
    \label{eq:green_fun_expand}
    G_{loc,a}(i\omega) \approx \frac{A_a}{i\omega} + \frac{B_a}{(i\omega)^2} + \frac{C_a}{(i\omega)^3},\quad 
    G_{x_i,x_j}'(i\omega)\approx \frac{A_{x_i,x_j}}{i\omega} + \frac{B_{x_i,x_j}}{(i\omega)^2} + \frac{C_{x_i,x_j}}{(i\omega)^3}
    \ea 
    where 
    \ba 
    & A_{x_i,x_j} = \frac{1}{N} \sum_{\kk,n=1,...,n_{flat} } U_{\kk,a_in}U_{\kk,a_jn}^*e^{i\kk\cdot(\RR_i-\RR_j+\bm{r}_{a_i} -\bm{r}_{a_j})},
    \quad A_{a_i} = A_{x_i,x_i}
    \nonumber\\ 
 &B_{x_i,x_j}=\frac{1}{N} \sum_{\kk,n=1,...,n_{flat}} \delta \epsilon_{\kk,n}U_{\kk,a_in}U_{\kk,a_jn}^*e^{i\kk\cdot(\RR_i-\RR_j+\bm{r}_{a_i} -\bm{r}_{a_j})} 
 ,\quad 
 B_{a_i} =  B_{x_i,x_i}
 \nonumber\\ 
 &C_{x_i,x_j}=\frac{1}{N} \sum_{\kk,n=1,...,n_{flat}} (\delta \epsilon_{\kk,n})^2U_{\kk,a_in}U_{\kk,a_jn}^*e^{i\kk\cdot(\RR_i-\RR_j+\bm{r}_{a_i} -\bm{r}_{a_j})} 
 ,\quad C_{a_i}=  C_{x_i,x_i}
 \label{eq:def_AB_fields}
    \ea 
    Since we have assumed all the electronic orbitals are equivalent, we also have 
    \ba 
    A_{a_i} = A, \quad B_{a_i}=B, \quad C_{a_i}=C 
    \ea 
\end{itemize}

Combining \cref{eq:green_fun_expand} and \cref{eq:J_xixj_from_Green}, we have 
\ba 
J_{x_i,x_j}
\approx & \frac{1}{\beta} \sum_{i\omega} \frac{\hubbard^2A^2}{4}
\frac{e^{i\omega0^+}}{ 
\bigg[ (i\omega)^2 - \frac{A^4\hubbard^2}{4}
\bigg]^2 }\nonumber\\
&
\bigg[ (i\omega)^2   |A_{x_i,x_j}|^2
+ |B_{x_i,x_j}|^2 + i\omega [A_{x_i,x_j}B_{x_j,x_i} + A_{x_j,x_i}B_{x_i,x_j}]
+A_{x_i,x_j}C_{x_j,x_i} +A_{x_j,x_i}C_{x_i,x_j}
\bigg]  \nonumber\\ 
&
+ \frac{1}{\beta}
\sum_{i\omega} 
\frac{\hubbard^2A^2}{4}
\frac{ A^3\hubbard^2 Be^{i\omega0^+} }{ \bigg[ (i\omega)^2 
- \frac{A^4\hubbard^2}{4}
\bigg]^3}
\bigg[i\omega  |A_{x_i,x_j}|^2  
+ A_{x_j,x_i}B_{x_i,x_j} + A_{x_i,x_j}B_{x_j,x_i}
\bigg] 
\nonumber\\
&+\frac{1}{\beta}\sum_{i\omega} \frac{\hubbard^2A^2}{4}\frac{
A^6\hubbard^4 (5B^2 -2 AC)
+ 4 A^2\hubbard^2(B^2+2AC)(i\omega)^2 
}{8\bigg[(i\omega)^2 - \frac{A^4\hubbard^2}{4}\bigg]^4}e^{i\omega0^+} |A_{x_i,x_j}|^2 
\ea 
We evaluate the Matsubara summation via contour integration in the zero-temperature limit with $\beta\rightarrow \infty $ limit (see \cref{sec:Matsubara_summation}). We then observe 

\ba 
J_{x_i,x_j}
\approx &
-\frac{\hubbard}{8} 
|A_{x_i,x_j}|^2 
+ \frac{1}{2A^4\hubbard} |B_{x_i,x_j}|^2 
\nonumber\\
&
+\frac{1}{2A^5\hubbard }
\bigg[
A\bigg( A_{x_i,x_j}C_{x_j,x_i} +A_{x_j,x_i}C_{x_i,x_j}\bigg) 
-3 B \bigg( A_{x_i,x_j}B_{x_j,x_i} +A_{x_j,x_i}B_{x_i,x_j} \bigg) 
- 3 (C-2B^2/A) |A_{x_i,x_j}|^2
\bigg] 
\ea

\subsection{Effective spin-spin action}
We can combine Eq.~\ref{eq:berry_phase} and Eq.~\ref{eq:seff_spin_spin} and obtain the following effective theory for the spin fields $n_{\xx}^\mu(\tau)$
\ba 
\label{eq:spin_spin_model_action}
S_{eff} = \sum_x iA[\bm{n}_x] + \int_\tau \sum_{x_i,x_j} J_{x_i,x_j}\bm{n}_{{x}_i}(\tau)\cdot \bm{n}_{{x}_j}(\tau) 
\ea 
with
\ba
A[\bm{n}_x] = &-\int_\tau \frac{\cos(\theta_x(\tau))+1}{2}\partial_\tau \phi_x(\tau)  \nonumber\\ 
J_{x_i,x_j} = &-\frac{\hubbard}{8} 
|A_{x_i,x_j}|^2 
+ \frac{1}{2A^4\hubbard} |B_{x_i,x_j}|^2 
\nonumber\\
&
+\frac{1}{2A^5\hubbard }
\bigg[
A\bigg( A_{x_i,x_j}C_{x_j,x_i} +A_{x_j,x_i}C_{x_i,x_j}\bigg) 
-3 B \bigg( A_{x_i,x_j}B_{x_j,x_i} +A_{x_j,x_i}B_{x_i,x_j} \bigg) 
- 3 (C-2B^2/A) |A_{x_i,x_j}|^2
\bigg]  \nonumber\\
 A_{x_i,x_j} =& \frac{1}{N} \sum_{\kk,n=1,...,n_{flat} } U_{\kk,a_in}U_{\kk,a_jn}^*e^{i\kk\cdot(\RR_i-\RR_j+\bm{r}_{a_i} -\bm{r}_{a_j})},  \quad \quad A= A_{x_i,x_i}\nonumber\\ 
 B_{x_i,x_j}=&\frac{1}{N} \sum_{\kk,n=1,...,n_{flat}}
 \delta \epsilon_{\kk,n}U_{\kk,a_in}U_{\kk,a_jn}^*e^{i\kk\cdot(\RR_i-\RR_j+\bm{r}_{a_i} -\bm{r}_{a_j})} 
 , \quad \quad B = B_{x_i,x_i}\nonumber\\ 
 C_{x_i,x_j}=&\frac{1}{N} \sum_{\kk,n=1,...,n_{flat}} (\delta \epsilon_{\kk,n})^2U_{\kk,a_in}U_{\kk,a_jn}^*e^{i\kk\cdot(\RR_i-\RR_j+\bm{r}_{a_i} -\bm{r}_{a_j})} 
 , \quad \quad  C= C_{x_i,x_i}
 \label{eq:def_parameters} 
\ea 
% and $\phi_0 = \frac{n_{flat}}{n_{sub}}$ at low-temperature limit (Eq.~\ref{eq:val_phi0}).
In general, the first term in $S_{eff}$ describes the Berry phase of the spin fields, reflecting the quantum character of the effective spin fields,
where the second term corresponds to the spin-spin interactions.

It is also useful to investigate the spin-spin coupling in the momentum space, which is defined as 
\begin{align}
    J_{ab}(\qq) =&\frac{1}{N}\sum_{\RR_i,\RR_j} J_{(\RR_i,a),(\RR_j,b)}
    e^{i\qq\cdot(\RR_j+\rr_{a} - \RR_i -\rr_{b})} 
    % (1-\delta_{\RR_j+\rr_a,\RR_i+\rr_b})
    % \nonumber\\
    % =& \frac{1}{N}\sum_\kk
    % \bigg\{ 
    % \bigg[ -\frac{U}{8} 
    % A_{\kk+\qq,ab}A_{\kk,ba}
    % \bigg]  
    % +\frac{1}{2A^4U}
    % \bigg[ A_{\kk+\qq,ab}C_{\kk,ba} + C_{\kk+\qq,ab}A_{\kk,ba} + B_{\kk+\qq,ab}B_{\kk,ba} 
    % \bigg]\bigg\} 
\end{align}
With only a single band near the Fermi energy, we find 
\begin{align}
\label{eq:spin_spin_coupling_one_band}
   J_{ab}(\qq)     =& \frac{1}{N}\sum_\kk 
    \bigg[
    \bigg( -\frac{\hubbard}{8} 
    -\frac{ 3C-2B^2/A}{2A^5\hubbard}
    \bigg) 
    + \frac{ (\delta \epsilon_{\kk,1})^2 +(\delta \epsilon_{\kk+\qq,1})^2 
    +\delta \epsilon_{\kk+\qq,1}\delta \epsilon_{\kk ,1}
    }{2A^4\hubbard}
    - 3B\frac{ \delta\epsilon_{\kk,1}
    +\delta \epsilon_{\kk+\qq,1}
    }{2A^5\hubbard}
    \bigg] \nonumber\\
    &
    U_{\kk+\qq,a1} U_{\kk+\qq,b1}^*
    U_{\kk,b1} U_{\kk,a1}^* 
\end{align}

\section{Effective spin model at the single-orbital atomic limit}
\label{sec:atomic}
We discuss our effective spin theory at the single-orbital limit. We consider a single atomic orbital at a square lattice with dispersions 
\ba 
\delta \epsilon_\kk = -2t(\cos(k_x) +\cos(k_y))
\ea 
We consider the limit of $|t|\ll \hubbard$ where we have a single narrow band near the Fermi energy. We have 
\ba 
 U_{\kk,11} = 1,\quad A =1 , \quad B=0,\quad C= 4t^2,\quad \quad \phi_{0} = 1 
\ea 
From Eq.~\ref{eq:def_parameters}, we find 
\ba 
&A_{\xx_i,\xx_j} = 0,\quad B_{\xx_i,\xx_j} = \frac{1}{N}\sum_{\kk} (-2t)(\cos(k_x)+\cos(k_y)) e^{i\kk\cdot(\RR_i-\RR_j)} = -t \sum_{\bm{e} \in \{(\pm 1,0),(0,\pm 1)\}} \delta_{\xx_i-\xx_j,\bm{e}} \nonumber\\ 
&C_{\xx_i,\xx_j}=0
\ea 
for $\xx_i \ne \xx_j$. 
Then the effective spin-spin interactions are 
\ba 
J_{\xx_i,\xx_j} = \frac{t^2}{2\hubbard}\sum_{\bm{e} \in \{(\pm 1,0),(0,\pm 1)\}} \delta_{\xx_i-\xx_j,\bm{e}}
\ea 
We thus conclude that our approach successfully recovers the conventional antiferromagnetic superexchange coupling.

\section{Effective spin model in the flat-band limit}
\label{sec:flat_band}
We study the flat-band limit, where the system develops an isolated flat band at the Fermi energy with
\ba 
\delta \epsilon_{\kk,n} = 0,\quad B=C=0
\ea 
 With $x_i \ne x_j$, we have 
\ba 
B_{x_i,x_j} = 0 ,\quad C_{x_i,x_j}=0
\ea 
and 
\ba 
J_{x_i,x_j }= -\frac{\hubbard}{8} A_{x_i,x_j}A_{x_j,x_i }
\label{eq:J_at_flat_band_limit}
\ea 
We further assume the flat-band is non-atomic such that the wavefunction or quantum geometry of the band generates a non-zero $A_{x_i,x_j} $. 
At this limit, since (from Eq.~\ref{eq:def_parameters}) 
\ba 
A_{x_i,x_j} = A_{x_j,x_i}^* , 
\ea 
the spin-spin interaction is purely ferromagnetic 
\ba 
\label{eq:QG_ferro}
J_{x_i,x_j} =-\frac{\hubbard}{8} A_{x_i,x_j}A_{x_j,x_i } = -\frac{\hubbard}{8} |A_{x_i,x_j}|^2 \le 0 
\ea 
Therefore, the ground state is ferromagnetic, which is consistent with the exact solution for the flat band system with finite quantum geometry. 

We also note that, in the atomic limit with a single orbital, we have  
\begin{equation}
U_{\kk, (a_i=1,n=1)} = 1,
\end{equation}
and consequently,
\begin{equation}
A_{x_i,x_j} = \frac{1}{N} \sum_{\kk} U_{\kk,(a_i=1,n=1)} U_{\kk,(a_j=1,n=1)}^* e^{i\kk\cdot(\RR_i - \RR_j)} = \delta_{\RR_i,\RR_j}.
\end{equation}
Since the spin-spin interaction is induced by $A_{x_i,x_j}$ with $x_i \ne x_j$, we find
\begin{equation}
J_{x_i,x_j} = 0, \quad \text{for }x_i \ne x_j.
\end{equation}
Therefore, for atomic orbitals that form an exactly flat band, the effective spin-spin coupling vanishes.

\subsection{Flat-band toy model and quantum-geometry-induced ferromagnetism }
To further illustrate how quantum geometry favors ferromagnetism, we consider a topological flat-band model inspired by the topological heavy-fermion (THF) model of twisted bilayer graphene (TBG)~\cite{PhysRevLett.129.047601}. The non-interacting part of the model reads
\ba
H_0 =
\sum_{\kk,\sigma,\alpha\gamma}
c_{\kk,\alpha,\sigma}
\begin{bmatrix}
0 & \gamma & 0\\ 
\gamma & 0 & v_\star (k_x-ik_y) \\ 
0 & v_\star(k_x+ik_y) &0
\end{bmatrix}_{\alpha\gamma}
c_{\kk,\gamma,\sigma}^\dag ,
\quad
c_{\kk,\alpha,\sigma}
=\begin{bmatrix}
f_{\kk, \sigma}^\dag & \tilde{c}_{\kk,1,\sigma} & \tilde{c}_{\kk,2,\sigma}
\end{bmatrix}_\alpha .
\ea
This model can be viewed as a simplified version of the THF model containing only a single valley and a single orbital degree of freedom. Here, $f_{\kk,\sigma}$ denotes the localized $f$ orbital, while $\tilde{c}_{\kk,1,\sigma}$ and $\tilde{c}_{\kk,2,\sigma}$ reside in momentum space and form a dispersive Dirac cone.
In the non-interacting limit, the system hosts a perfectly flat band with nontrivial quantum geometry at the Fermi energy. In addition, the localized $f$ orbital experiences a strong on-site Hubbard interaction
$
\mathcal{U}a = U\delta_{a,1}$. 
The long-range order favored by the flat bands in TBG has been extensively discussed~\cite{KAN19,BER21a,BER21b}. In particular, ferromagnetism has been established in the absence of lattice relaxation and strain effects. 
Here, we use the present model to illustrate how quantum geometry can generate ferromagnetic correlations through a mechanism different from that proposed in Ref.~\cite{PhysRev.115.2}. 
The first key distinction is that the flat band here does not admit a single symmetric Wannier-orbital representation due to its nontrivial topology, already placing the system beyond the scope of Ref.\cite{PhysRev.115.2}. 
Although a large portion of the flat-band weight originates from localized Wannier-like orbitals, denoted here as $f$ electrons, the direct exchange coupling between these localized orbitals, which has been discussed in Ref.\cite{PhysRev.115.2} as the origin of ferromagnetism, is negligibly small and is therefore typically neglected.
Nevertheless, even in the absence of direct exchange coupling, the nontrivial quantum geometry generated by the hybridization between the $f$ and $\tilde{c}$ electrons gives rise to a strong ferromagnetic tendency. In this sense, the ferromagnetic interaction originates purely from the quantum geometry, or equivalently, the nontrivial wavefunction structure of the flat band. Using Eq.~\cref{eq:QG_ferro}, the resulting quantum-geometry-induced ferromagnetic coupling between localized $f$ orbitals can be directly evaluated as $\sim \frac{U}{8N}\sum_\kk
\frac{|v_\star \kk|^2}{\gamma^2+|v_\star\kk|^2}
\cos(\kk\cdot\RR)$.

\section{FM-AFM transition} 
\label{sec:fm_afm}
We study the energy competition between the FM and AFM states. We focus on the case where a single-flat band appears near the Fermi-energy, leading to the effective spin-spin coupling given in \cref{eq:spin_spin_coupling_one_band}. We observe that in the classical limit (where we have ignored the $\tau$-dependency of $\b,{n}_{x_i}$ fields), the energy of the system and then be effectively written as 
\begin{align}
    E =\sum_{x_i,x_j} J_{x_i,x_j} \bm{n}_{x_i}\cdot \bm{n}_{x_j}
\end{align}
with $\bm{n}_{x_i}$ the unit vector characterizing the spin orientation. 
The configuration $\bm{n}_{x_i}$ that minimizes the energy $E$ will be the ground state energy in the classical limit. It is then also useful to introduce the momentum space formula
\begin{align}
    \bm{n}_{x_i = (\RR_i,a_i)} = \frac{1}{\sqrt{N}}\sum_\qq \bm{n}_{\qq,a_i}e^{i\qq\cdot(\RR_i+\rr_{a_i})}
\end{align}
We have 
\begin{align}
    E = \sum_{\qq,ab} J_{ab}(\qq) \bm{n}_{-\qq,a}\cdot \bm{n}_{\qq,b}
\end{align}
where the momentum-space coupling is defined as 
\begin{align}
    J_{a_ib_j}(\qq) = \frac{1}{N} \sum_{(\RR_i,a_i),(\RR_j,a_j)}J_{(\RR_i,a_i),(\RR_j,a_j)}
    e^{i\qq\cdot(\RR_j+\rr_{a_j}-\RR_i-\rr_{a_i})}
\end{align}

To study the magnetic order of the system, it is useful to study the eigenvalue of the matrix $J_{a_i,b_j}(\qq)$. The momentum where the smallest eigenvalue is realized corresponds to the magnetic order that the interaction favors.

We first investigate the flat-band limit, where $\delta \epsilon_{\kk,1}= 0 $. We have (see \cref{eq:def_parameters}) 
\begin{align}
    J_{x_ix_j} = -\frac{\hubbard}{8} |A_{x_i,x_j} A_{x_j,x_i}|^2  \le 0 
\end{align}
We observe that $J_{x_i,x_j}$ is always ferromagnetic ($\qq=0$ order). 
% Therefore, the spin-configuration that minimizes energy in such a limit. 
We then investigate the spin-spin coupling matrix at $\qq=0$. We find 
\begin{align}
    J_{a b}(\qq=0) = -\frac{\hubbard}{8}\frac{1}{N}\sum_\kk |U_{\kk,a 1}|^2 |U_{\kk,b 1}|^2 
\end{align}
Since all the spin-spin couplings are ferromagnetic, the eigenvector of $J_{ab}(\qq=0)$ with the lowest eigenvalue is $[v]_a = \frac{1}{\sqrt{n_{sub}}}$ with $n_{sub}$ the number of sublattices.  
The corresponding eigenvalue is 
\begin{align}
    E_{\qq=0, lowest} = -\frac{\hubbard}{8n_{sub}}\frac{1}{N}\sum_{\kk,ab} |U_{\kk,a1}|^2 |U_{\kk,b1}|^2 = -\frac{\hubbard}{8n_{sub}} = -\frac{\hubbard A}{8}
\end{align}

We now discuss the effect of finite dispersion. 
When the band dispersion is finite ($\delta \epsilon_{\kk,1} \ne 0$), the spin-spin couplings are no longer purely ferromagnetic. The emergence of antiferromagnetic coupling implies that the lowest eigenvalue of $J_{ab}(\qq)$ may occur at a finite momentum $\qq$, indicating a tendency toward antiferromagnetic ordering. 
To investigate this potential instability, it is useful to expand $E_{\qq,\text{lowest}}$ in powers of $\qq$ around $\qq = 0$, in the presence of finite dispersion. 

To perform such expansion, we first consider the effect of finite dispersion at $\qq=0$.
From \cref{eq:spin_spin_coupling_one_band}, we have 
\begin{align}
     J_{a b}(\qq=0) =\frac{1}{N}\sum_\kk \hubbard
     \bigg[ \frac{-1}{8} + \alpha_\kk 
     \bigg] 
    |U_{\kk,a 1}|^2 |U_{\kk,b 1}|^2 
\end{align}
where 
\ba 
\label{eq:def_additional}
\alpha_\kk = 
\frac{3(\delta \epsilon_{\kk,1})^2}{2A^4\hubbard^2} 
-\frac{3C-2B^2/A}{2A^5\hubbard^2}
-\frac{3B \delta \epsilon_{\kk,1}}{A^5\hubbard^2}
\ea 
Since $C \sim \delta \epsilon_\kk^2, B \sim \delta\epsilon_\kk$ (\cref{eq:def_parameters}), we note that $\alpha_\kk \sim \delta \epsilon_\kk^2/\hubbard^2 $. 

% Since we consider a flat-band limit where $|\delta \epsilon_{\kk,1}|/\hubbard \ll 1 $, we expect 
% \ba 
% \frac{\hubbard}{8}\gg \frac{ 3 (\delta \epsilon_{\kk,1})^2
%     }{2A^4\hubbard}
% \ea 
% and then 
% \ba 
%  J_{a b}(\qq=0) \approx \frac{1}{N}\sum_\kk 
%      \bigg[ \frac{-\hubbard}{8}
%      \bigg] 
%     |U_{\kk,a 1}|^2 |U_{\kk,b 1}|^2 
% \ea 
As long as we remain in the narrow-band limit where $|\delta \epsilon_\kk|/\hubbard \ll 1$, we expect the wavefunction associated with the lowest eigenvalue of $J_{ab}(\qq=0)$ remain 
\ba 
\label{eq:egv_at_q=0}
[v]_a = \frac{1}{\sqrt{n_{sub}}}
\ea 
% with 
% \ba 
% E_{\qq=0,lowest} = \sum_{ab}[v]_a J_{ab}(\qq=0) [v]_b
% \ea 

We now examine $E_{\qq,\text{lowest}}$ at finite $\qq$. 
% We next discuss the behaviors of the lowest eigenvalue away from $\qq=0$ point. A
At small $\qq$, we project to the eigenstates at $\qq=0$ (\cref{eq:egv_at_q=0}) and find 
\begin{align}
\label{eq:Eq_lowest_expansion}
    E_{\qq,lowest}  \approx  &
    \sum_{a,b} [v^*]_a J_{ab}(\qq) [v]_b \nonumber\\ 
     \approx &
      E_{\qq=0,lowest}
      + \sum_\mu \sum_{a,b} \frac{1}{n_{sub}}q^\mu \partial_{q^\mu}J_{ab}(\qq)
      \bigg|_{\qq=0} \nonumber\\ 
      & + \frac{1}{2}\sum_{\mu\nu} 
       \sum_{a,b}  \frac{1}{n_{sub}} q^\mu q^\nu \partial_{q^\mu}\partial_{q^\nu}J_{ab}(\qq)
      \bigg|_{\qq=0} 
\end{align}
We now show that the first-order term vanishes
\begin{align}
\label{eq:Jq_first_order_term}
   & \sum_{a,b}   \partial_{q^\mu}J_{ab}(\qq)\nonumber\\
    =& \frac{1}{N}
    \sum_{\kk,ab} 
    \bigg[-\frac{\hubbard}{8} 
    +\hubbard \alpha_\kk 
    \bigg] \bigg[ \partial_{k^\mu} U_{\kk,a1}U_{\kk,b1}^* U_{\kk,b1}U_{\kk,a1}^* + 
   U_{\kk,a1} \partial_{k^\mu} U_{\kk,b1}^* U_{\kk,b1}U_{\kk,a1}^*
    \bigg]  \nonumber\\ 
& + 
\frac{1}{N}
    \sum_{\kk,ab} 
    \bigg[ 
    +\frac{ 3(\delta \epsilon_{\kk,1}) \partial_{k^\mu}\epsilon_{\kk,1} 
    - 3B\partial_{k^\mu}\delta \epsilon_{\kk,1}A 
    }{2A^4\hubbard}
    \bigg] |U_{\kk,a1}|^2 |U_{\kk,b1}|^2 \nonumber\\ 
    =& + \frac{1}{N}
    \sum_{\kk,ab} 
    \bigg[ 
    +\frac{ 3(\delta \epsilon_{\kk,1}) \partial_{k^\mu}\delta \epsilon_{\kk,1} 
    - 3B\partial_{k^\mu}\delta \epsilon_{\kk,1}A 
    }{2A^4\hubbard}
    \bigg] |U_{\kk,a1}|^2 |U_{\kk,b1}|^2\nonumber\\ 
= &\frac{1}{N}
    \sum_{\kk} 
    \bigg[\frac{ 3(\delta \epsilon_{\kk,1}) \partial_{k^\mu}\delta \epsilon_{\kk,1} 
    - 3B\partial_{k^\mu}\delta \epsilon_{\kk,1}A 
    }{2A^4\hubbard}
    \bigg]  \nonumber\\ 
= &\frac{1}{N}\sum_{\kk}\frac{ 6(\delta \epsilon_{\kk,1}) \partial_{k^\mu}\epsilon_{\kk,1}}{2A^4\hubbard }
\end{align}
We notice that for an isolated flat band, the dispersion $\delta \epsilon_{\kk,1}$ is an analytical function of $\kk$ with the following periodicity
\ba 
\delta \epsilon_{\kk+\sum_\mu n^\mu \bm{b}_\mu,1} =\delta \epsilon_{\kk,1 },\quad n^\mu \in \mathbb{Z}
\ea 
where $\{\bm{b}_\mu\}_\mu $ are the reciprocal lattice vectors. Therefore, we conclude 
\ba 
\label{eq:translational_symmetry_dispersion}
\sum_{\kk'} (\delta \epsilon_{\kk' + \kk ,1})^n = \sum_{\kk'} (\delta \epsilon_{\kk',1} )^n,\quad n \in \{1,2\}
\ea 
and then 
\ba 
\label{eq:label_zero_of_de_e}
&0 = \partial_{k^\mu} \sum_{\kk'}( \delta \epsilon_{\kk' + \kk,1 })^n ] |_{\kk =0 }
\nonumber\\ 
\Rightarrow &0= \sum_{\kk'}  2 \delta \epsilon_{\kk',1} \partial_{k'^\nu} \delta \epsilon_{\kk',1} ,\quad \text{and},\quad 
0= \sum_{\kk'}   \partial_{k'^\nu} \delta \epsilon_{\kk',1} 
\ea 
Therefore, combining \cref{eq:Jq_first_order_term,eq:label_zero_of_de_e}, we conclude 
\begin{align}
     \sum_{a,b} \partial_{q^\mu}J_{ab}(\qq)=0 
\end{align}

We next consider the second-order contributions of  \cref{eq:Eq_lowest_expansion}. 
We find 
\begin{align}
   \frac{1}{2} \sum_{ab} \partial_{q^\mu} \partial_{q^\nu}J_{ab}(\qq)\bigg|_{\qq=0}= &
    \frac{\hubbard }{N}\sum_{\kk,\mu\nu } 
(-\sum_a \partial_{k^\mu} U_{\kk,a1}\partial_{k^\nu} U_{\kk,a1}^*
+ \sum_{a,b} \partial_{k^\mu}U_{\kk,a1}\partial_{k^\nu}U_{\kk,b1}^* U_{\kk,b1}U_{\kk,a1}^*) 
(-\frac{1}{8} +\alpha_\kk ) \nonumber\\ 
&-\frac{1}{2A^4\hubbard N } \sum_{\kk,\mu\nu }\frac{\partial_{k^\mu}\delta \epsilon_{\kk, 1} \partial_{k^\nu}\delta\epsilon_{\kk,1} 
}{2}
\end{align}
where we use \cref{eq:translational_symmetry_dispersion} and obtain 
\ba 
&0 = [\partial_{k^\mu}\partial_{k^\nu} \sum_{\kk'}\delta \epsilon_{\kk' + \kk,1 }^2 ] |_{k^\mu =0 }
\nonumber\\ 
\Rightarrow &0= \sum_{\kk'}  2 \partial_{k'^\mu} \delta \epsilon_{\kk',1} \partial_{k'^\nu} \delta \epsilon_{\kk',1} +  2 \partial_{k'^\mu} \partial_{k'^\nu}\delta \epsilon_{\kk',1}  \delta \epsilon_{\kk',1}
\ea 
We then find 
\begin{align}
     \frac{1}{2} \sum_{ab} \partial_{q^\mu} \partial_{q^\nu}J_{ab}(\qq)\bigg|_{\qq=0}= 
    &+\frac{\hubbard }{N}\sum_{\kk,\mu\nu } 
(-\sum_a \partial_{k^\mu} U_{\kk,a1}\partial_{k^\nu} U_{\kk,a1}^*
+ \sum_{a,b} \partial_{k^\mu}U_{\kk,a1}\partial_{k^\nu}U_{\kk,b1}^* U_{\kk,b1}U_{\kk,a1}^*) 
(-\frac{1}{8} + \alpha_\kk ) \nonumber\\ 
&-\frac{1}{4A^4\hubbard N } \sum_{\kk,\mu\nu }\partial_{k^\mu}\delta \epsilon_{\kk, 1} \partial_{k^\nu}\delta\epsilon_{\kk,1} 
\end{align}
% We observe 
% \begin{align}
%     \frac{1}{2}  \sum_{ab} \partial_{q^\mu} \partial_{q^\nu}J_{ab}(\qq)\bigg|_{\qq=0}= 
%     &+\frac{\hubbard }{N}\sum_{\kk,\mu\nu } 
% (-\sum_a \partial_{k^\mu} U_{\kk,a1}\partial_{k^\nu} U_{\kk,a1}^*
% + \sum_{a,b} \partial_{k^\mu}U_{\kk,a1}\partial_{k^\nu}U_{\kk,b1}^* U_{\kk,b1}U_{\kk,a1}^*) 
% (-\frac{1}{8} +\frac{3\delta \epsilon_{\kk}^2}{2A^4\hubbard ^2} ) \nonumber\\ 
% &-\frac{1}{4A^4\hubbard N } \sum_{\kk,\mu\nu }\partial_{k^\mu}\delta \epsilon_{\kk, 1} \partial_{k^\nu}\delta\epsilon_{\kk,1} 
% \end{align}
We introduce 
\ba 
\label{eq:def_mass_Q_matrix}
&Q_{\mu\nu}(\kk) = \sum_{a,b}  \partial_{k^\mu} U_{\kk,a}^* 
\bigg( \delta_{a,b} - U_{\kk,a} U_{\kk,b}^*\bigg)
\partial_{k^\nu}U_{\kk,b}
,\quad\quad Q = \frac{1}{N}\sum_\kk Q_{\mu\nu}(\kk) 
\nonumber\\ 
&m_{\mu\nu}(\kk) = \partial_{k^\mu } \delta \epsilon_{\kk,1} \partial_{k^\nu}\delta \epsilon_{\kk,1} ,\quad\quad M_{\mu\nu} = \frac{1}{N}\sum_\kk m_{\mu\nu}(\kk) 
\ea 
and obtain 
\ba 
\frac{1}{2}  \sum_{ab} \partial_{q^\mu} \partial_{q^\nu}J_{ab}(\qq)\bigg|_{\qq=0}
= 
\hubbard \frac{1}{N}\sum_\kk  \bigg\{Q_{\mu\nu}(\kk)
\bigg( \frac{1}{8} - \alpha_\kk \bigg) 
    -\frac{m_{\mu\nu}(\kk)}{4A^4\hubbard ^2}
    \bigg\} 
\ea 
In practice, we assume that $Q_{\mu\nu}(\kk)$ and $ |\delta \epsilon_{\kk,1}|^2/\hubbard^2$ are of the same order, so that the effects of quantum geometry and band dispersion can compete. 
Consequently, the term $Q_{\mu\nu}(\kk) \alpha_\kk $ with $\alpha_\kk \sim |\delta \epsilon_\kk|^2/\hubbard^2$ is expected to be much smaller.
By dropping $Q_{\mu\nu}(\kk) \alpha_\kk $ term, we approximately have
\begin{align}
\label{eq:Jq_exp}
    \frac{1}{2}  \sum_{ab} \partial_{q^\mu} \partial_{q^\nu}J_{ab}(\qq)\bigg|_{\qq=0}\approx \hubbard  \bigg\{\frac{Q_{\mu\nu}}{8}
    -\frac{M_{\mu\nu}}{4A^4\hubbard ^2}
    \bigg\} 
\end{align}
Then we finally have (from \cref{eq:Eq_lowest_expansion,eq:Jq_exp}
\begin{align}
\label{eq:Eq_expansion_small_q}
  E_{lowest,\qq} \approx &  E_{lowest,\qq=0} 
    \nonumber\\
    & +\frac{1}{n_{sub}}\hubbard  \sum_{\mu\nu} \bigg\{\frac{Q_{\mu\nu}}{8}
    -\frac{M_{\mu\nu}}{4A^4\hubbard ^2}
    \bigg\} q^\mu q^\nu 
\end{align}
Therefore, $E_{lowest,\qq}$ no longer reach its minimum at $\qq=0$ when the determinant of the matrix
\begin{align}
H_{\mu\nu} = A\hubbard \bigg[ \frac{Q_{\mu\nu}}{8}
    -\frac{M_{\mu\nu}}{4A^4\hubbard ^2}\bigg] 
\end{align}
becomes negative. This then indicates the instability of the FM phase. 

% Here, we also demonstrate that, for a 2D system with $C_{4z}$, $C_{3z}$ or $C_{6z}$ symmetry, $H_{\mu\nu}$ is proportional to identity matrix. Without loss of generality, we consider the effect of $C_{nz}$ symmetry (where $n\in \{4,3,6\}$).
% In addition, since we have assumed the system has $SU(2)$ symmetry, we consider spinless rotation. 
% Under $C_{nz}$ rotation, we have
% \begin{align}
%     C_{nz}\bm{S}_{\qq, a} C_{nz}^{-1} = \sum_b \bm{S}_{C_{nz}\qq,b}
%     D[C_{nz}]_{ba}
% \end{align}
% where $D[C_{nz}]_{ba}$ are the representation matrix. Then $J_{ab}(\qq)$ follows 
% \begin{align}
%     J_{ab}(C_{3z}\qq) =\sum_{c,d} D[C_{nz}]_{ac}J_{cd}(\qq)D[C_{nz}]_{bd}
% \end{align}
% Then, this leads to 
% \begin{align}
%    \sum_{\mu\nu} H_{\mu\nu} [C_{3z}\qq]^\mu [C_{3z}\qq]^\nu 
%    =\sum_{a,c,b,d,\mu\nu} [v]_a D[C_{nz}]_{ac}[v]_c
%    [v]_b D[C_{nz}]_{ac}[v]_d  H_{\mu\nu} q^\mu q^\nu 
% \end{align}

% The transition point is approximately 
% \begin{align}
% \label{eq:transition_fm_afm}
%     Q = \frac{2M}{A^4U^2} 
% \end{align}
% where $Q = \sum_{\mu}Q_{\mu\mu}, M_{\mu\nu} = \sum_\mu M_{\mu\nu}$. 

\section{Spin stiffness}
In this section, we also provide detailed calculations of the spin stiffness of the ferromagnetic state to check the stability of the flat-band ferromagnetism\cite{peotta_superfluidity_2015,julku_geometric_2016,torma_superconductivity_2022,liang_band_2017,huhtinen_revisiting_2022,herzog-arbeitman_superfluid_2022,herzog-arbeitman_many-body_2022,yu2025quantumgeometryquantummaterials}. 
For convenience, we consider a simple situation with only one flat band appearing near the Fermi energy, and all the sublattices are equivalent/symmetry-related. 
In this section, we provide the calculation of stiffness from the original action given in \cref{eq:eff_action_before_expand}. 
We also note that the spin model derived in \cref{eq:spin_spin_model_action} captures only two-body interaction terms, neglecting higher-order spin interactions such as $n^{\mu}{x_i} n^{\nu}{x_j} n^{\gamma}_{x_m}$. As a result, it cannot fully reproduce the exact spin stiffness of the ferromagnetic state. 
To recover the exact result, one can perform the calculation using the exact action presented in \cref{eq:eff_action_before_expand}.

The exact action (\cref{eq:eff_action_before_expand}) takes the form of
\begin{align}
\label{eq:eff_action_before_expand_2}
S= S_\phi -
\text{Tr}[\log(\tilde{G}^{-1} + V) ] 
\end{align}
At low-temperature limits, we have
\begin{align}
    \bm{\phi}_{x_i} = \phi_{0} \bm{n}_{x_i},\quad \phi_{0}= A
\end{align}
Since all the sublattices are equivalent, we also have
\begin{align}
    A= A_a = \frac{1}{N}\sum_{\kk}|U_{\kk,a1}|^2 =\frac{1}{n_{sub}}
\end{align}

We use the Holstein-Primakoff boson to calculate the stiffness of the ferromagnetic state 
\ba 
{n}^z_{x_i} = 1- 2a_{x_i}^\dag a_{x_i}
,\quad n^x_{x_i}= (a_{x_i}+a_{x_i}^\dag) ,\quad n^y_{x_i} = \frac{1}{i}(a_{x_i}-a_{x_i}^\dag) 
\label{eq:HP_boson}
\ea 
Then we could expand the action $S$ to a quadratic action of the $a,a^\dag$ fields which describe the spin fluctuations of the system. 
We first separate interaction vertex $V$ (\cref{eq:gree_fun_full}) into three parts
\begin{align}
    &V= v_0+\delta v_1 + \delta v_2 \nonumber\\ 
    &[v_0]_{(x_i,\sigma,i\omega),(x_j,\sigma',i\omega')} = \delta_{x_i,x_j}\frac{\hubbard  \phi_0 }{2\beta}\sigma^z_{\sigma,\sigma'}e^{i\omega' 0^+}\nonumber\\
    &[\delta v_1]_{(x_i,\sigma,i\omega),(x_j,\sigma',i\omega')} = \delta_{x_i,x_j}
    \frac{\hubbard  {A} }{\beta}e^{i\omega' 0^+}
    \begin{bmatrix}
        0 & a_{x_i}^\dag (-i\omega+i\omega') \\
        a_{x_i}(i\omega-i\omega') & 0
    \end{bmatrix}_{\sigma,\sigma'} \nonumber\\ 
    &[\delta v_2]_{(x_i,\sigma,i\omega),(x_j,\sigma',i\omega')} =- \delta_{x_i,x_j}\sigma \delta_{\sigma,\sigma'}
    \frac{\hubbard  {A} }{\beta}
    \frac{1}{\beta}\sum_{i\omega''}a_{x_i}^\dag(i\omega'') a_{x_i}(i\omega-i\omega'+i\omega'')
    % (a_{x_i}^\dag a_{x_i})(i\omega-i\omega') \sigma^z e^{i\omega' 0^+}
\end{align} 
where we combine \cref{eq:gree_fun_full,eq:HP_boson} and expand $V$ in powers of $a,a^\dag$ fields. In addition, the operator in Matsubara frequency is defined as
\ba 
a_{x_i}(i\omega) = \int_\tau a_{x_i}(\tau) e^{i\omega\tau}
\ea 

Since $S_{\phi}$ does not depend on $\bm{n}_{x_i}$, only the trace part in \cref{eq:eff_action_before_expand_2} contributes to the spin stiffness and can be written as 
\begin{align}
    S_{stiff} = -\text{Tr}[\log(\tilde{G}^{-1} +v_0 + \delta v_1  + \delta v_2)]
\end{align}
We expand the above action to quadratic order in $a,a^\dag$ and find 
\begin{align}
\label{eq:stiffness_expand}
    S_{stiff} \approx S_{const}  
    -\text{Tr}[ (\tilde{G}^{-1}+v_0)^{-1} \delta v_2 ] + \frac{1}{2}
    \text{Tr}[ (\tilde{G}^{-1}+v_0)^{-1} \delta v_1 (\tilde{G}^{-1}+v_0)^{-1} \delta v_1 ] 
\end{align}
where $S_{const}$ denotes the term that does not depend on $a,a^\dag$ fields. 
We evaluate the remaining two terms. For later convenience, we also introduce 
\begin{align}
    G_v =  [\tilde{G}^{-1}+v_0]^{-1}\, .
\end{align}
We find 
\begin{align}
  [G_v]_{(\xx_i,\sigma,i\omega), (\xx_j,\sigma',i\omega')} = \delta_{i\sigma,i\omega'}\delta_{\sigma,\sigma'} \frac{1}{N}\sum_\kk \bigg[ i\omega \mathbb{I} - t_\kk + \frac{\hubbard {A}}{2}\sigma \mathbb{I}\bigg]^{-1}_{a_i,a_j} e^{i\kk\cdot(\xx_i-\xx_j)}  
\end{align}
We focus on the contributions from the flat band. By projecting to the flat band, we have 
\begin{align}
   & [G_v]_{(\xx_i,\sigma,i\omega), (\xx_j,\sigma',i\omega')} \approx \delta_{i\omega,i\omega'}\delta_{\sigma,\sigma'}e^{i\omega' 0^+}\frac{1}{{N}}\sum_\kk \frac{1}{i\omega - \delta \epsilon_{\kk,1} + \sigma \frac{\hubbard {A}}{2} }U_{\kk,a_i1}U_{\kk,a_j1}^* e^{i\kk\cdot(\xx_i-\xx_j)} 
\end{align}

We now evaluate $S_{stiff}$. 
The second term in the $S_{stiff}$ (\cref{eq:stiffness_expand}) reads
\begin{align}
& -\text{Tr}[G_v \delta v_2 ] \nonumber\\
    =&\sum_{x_i,i\omega,\sigma}\frac{\hubbard \phi_{0}}{\beta}[G_v]_{(x_i,\sigma,i\omega),(x_i,\sigma,i\omega)} 
   \frac{\sigma}{\beta}\sum_{i\omega'} a_{x_i}^\dag(i\omega') a_{x_i}(i\omega')
    e^{i\omega 0^+} \nonumber\\ 
    =& \frac{1}{\beta}\sum_{x_i,i\omega,\sigma}\hubbard {A} a_{x_i}^\dag (i\omega) a_{x_i}(i\omega) \frac{1}{N}\sum_\kk |U_{\kk,a_i1}|^2\frac{\sigma+1}{2} \nonumber\\ 
    =& \frac{1}{\beta }\sum_{x_i,i\omega}
    \hubbard  {A}^2 a_{x_i}^\dag (i\omega) a_{x_i}(i\omega) 
\end{align}
The third term in the $S_{stiff} $(\cref{eq:stiffness_expand}) reads
\begin{align}
    & \frac{1}{2}\text{Tr}[G_v \delta v_1 G_v \delta v_1] \nonumber\\ 
    =&\frac{1}{2}\sum_{i\omega_1,i\omega_2, x_i,x_j,\sigma_1,i\sigma_2} 
    \frac{\hubbard ^2{A}^2}{\beta^2N^2}\sum_{\kk_1,\kk_2}
    \frac{U_{\kk_1,a_i1}U_{\kk_1,a_j 1}^*}{i\omega_1 - \delta \epsilon_{\kk_1,1}+\sigma_1 \hubbard {A}/2}
     \frac{ U_{\kk_2,a_j1}U_{\kk_2,a_i1}^*}{i\omega_2 - \delta \epsilon_{\kk_2,1}+\sigma_2 \hubbard {A}/2}e^{i(\kk_1-\kk_2)(\xx_i-\xx_j)} \nonumber\\ 
     &\begin{bmatrix}
         & a_{x_j}^\dag(-i\omega_1+i\omega_2) \\ 
         a_{x_j}(i\omega_1-i\omega_2)
     \end{bmatrix}_{\sigma_1,\sigma_2}
     \begin{bmatrix}
         & a_{x_i}^\dag(-i\omega_2+i\omega_1) \\ 
         a_{x_i}(i\omega_2-i\omega_1)
     \end{bmatrix}_{\sigma_2,\sigma_1} \nonumber\\ 
     =&\frac{\hubbard ^2{A}^2}{2\beta N^2}\sum_{x_i,x_j}\sum_{i\Omega} \sum_{\kk_1,\kk_2} U_{\kk_1,a_i1}U_{\kk_1,a_j1}^* U_{\kk_2,a_j 1}U_{\kk_2,a_i1}^* e^{i(\kk_1-\kk_2)(\xx_i-\xx_j)} \nonumber\\
     &\bigg[\frac{1}{i\Omega - \delta \epsilon_{\kk_1,1} +\delta \epsilon_{\kk_2,1} -\hubbard {A}}a_{\xx_j}(i\Omega)a_{\xx_i}^\dag(i\Omega) 
     +\frac{1}{i\Omega - \delta_{\kk_2,1} +\delta \epsilon_{\kk_1,1}-\hubbard {A}}a_{\xx_j}^\dag(i\Omega) a_{\xx_i}(i\Omega)\bigg]  \nonumber\\ 
     \approx & 
     \frac{\hubbard ^2{A}^2}{2\beta N^2}\sum_{x_i,x_j}\sum_{i\Omega} \sum_{\kk_1,\kk_2} U_{\kk_1,a_i1}U_{\kk_1,a_j1}^* U_{\kk_2,a_j 1}U_{\kk_2,a_i1}^* e^{i(\kk_1-\kk_2)(\xx_i-\xx_j)} \nonumber\\
     &\bigg[-\frac{1 + 
     \frac{i\Omega}{\hubbard {A}-\delta \epsilon_{\kk_1,1}+\delta \epsilon_{\kk_2,1} }
     }{\hubbard {A}-\delta \epsilon_{\kk_1,1}+\delta \epsilon_{\kk_2,1}}a_{\xx_j}(i\Omega)a_{\xx_i}^\dag(i\Omega) -
     \frac{1 
     +\frac{i\Omega}{\hubbard {A}+\delta \epsilon_{\kk_1,1}-\delta \epsilon_{\kk_2,1}}
     }{\hubbard {A}+\delta \epsilon_{\kk_1,1}-\delta \epsilon_{\kk_2,1}}a_{\xx_j}^\dag(i\Omega)a_{\xx_i}(i\Omega)
     \bigg] 
\end{align}
where in the final line we have expanded in powers of frequency $\Omega$. 
Since we are interested in the low-energy behavior, corresponding to the low-frequency regime, we retain terms up to first order in $i\Omega$. Higher-order terms, while in principle capable of further renormalizing the dispersion of the $a$ fields, are expected to be less relevant in the low-energy limit.

It is useful to work in the momentum space. We perform the following Fourier transformation with 
\begin{align}
    a_{\xx_i}(i\omega) = \frac{1}{\sqrt{N}}\sum_\kk a_{\kk,a_i}(i\omega)e^{i\kk\cdot \xx_i }
\end{align}
Then the low-energy effective theory of $a,a^\dag$ reads
\begin{align}
    S_{stiff} \approx &
    \frac{1}{\beta} \sum_{i\Omega,\qq, a,a'}
    \bigg[ \hubbard {A}^2 \delta_{a,a'}
    \nonumber\\ 
&
-\frac{1}{N}
\sum_\kk U_{\kk+\qq,a 1}U_{\kk+\qq,a'1}^* U_{\kk,a'1}U_{\kk,a1}^*\frac{\hubbard ^2{A}^2}{\hubbard {A} - \delta \epsilon_{\kk+\qq,1} +\delta\epsilon_{\kk,1}}
\bigg( 1+ \frac{i\Omega}{\hubbard {A} - \delta \epsilon_{\kk+\qq,1} +\delta\epsilon_{\kk,1}}
\bigg)
    \bigg] a_{\qq,a}^\dag (i\Omega) a_{\qq,a'}(i\Omega) \nonumber\\ 
    =& \frac{1}{\beta}\sum_{i\Omega,\qq}
    \bigg[ M_{\qq,aa'} -N_{\qq,aa'}i\Omega\bigg]a_{\qq,a}^\dag(i\Omega) a_{\qq,a'}(i\Omega)
\end{align}
where 
\begin{align}
    &N_{\qq,aa'} = \frac{1}{N}\sum_\kk 
    \frac{\hubbard ^2{A}^2}{(\hubbard {A} 
    -\delta \epsilon_{\kk+\qq,1}+\delta \epsilon_{\kk,1})^2
    }U_{\kk+\qq,a1}U_{\kk+\qq,a'1}^* U_{\kk,a'1}U_{\kk,a1}^* \nonumber\\ 
    &M_{\qq,aa'}={\hubbard {A}^2}{}\delta_{a,a'} -
     \frac{1}{N}\sum_\kk 
    \frac{\hubbard ^2{A}^2}{\hubbard {A} 
    -\delta \epsilon_{\kk+\qq,1}+\delta \epsilon_{\kk,1}
    }U_{\kk+\qq,a1}U_{\kk+\qq,a'1}^* U_{\kk,a'1}U_{\kk,a1}^*
\end{align}

We now investigate the gapless mode.
The Goldstone mode can be written as 
\begin{align}
    A_{\qq}(i\Omega) = \sum_a \frac{1}{\sqrt{n_{sub}}}a_{\qq,a}^\dag(i\Omega)
\end{align}
 As we show later, $A_{\qq}(i\Omega)$ develops a gapless mode at $\qq=0$. Projecting to the Goldstone mode, we take $a_{\qq,a}^\dag(i\Omega) \approx \frac{1}{\sqrt{n_{sub}}}A_\qq(i\Omega)$. The effective theory of Goldstone mode reads
\begin{align}
    S_{sitff, G} \approx & \frac{1}{\beta}\sum_{i\Omega,\qq}\frac{1}{n_{sub}}\sum_{aa'}\bigg[ M_{\qq,aa'}-N_{\qq,aa'}i\Omega\bigg] A^\dag_{\qq}(i\Omega)A_{\qq}(i\Omega) \nonumber\\ 
    =& 
    \frac{1}{\beta}\sum_{i\Omega,\qq}\frac{N_\qq}{n_{sub}}\bigg[ \frac{M_\qq}{N_\qq}-i\Omega\bigg] A^\dag_{\qq}(i\Omega)A_{\qq}(i\Omega)
\end{align}
where we have introduced $M_\qq = \sum_{aa'}M_{\qq,aa'},N_\qq =\sum_{aa'}N_{\qq,aa'}$. 
Therefore, the Green's function of Goldstone mode is proportional to $1/(i\Omega -M_\qq/N_\qq)$. Then the dispersion of Goldstone mode is 
\begin{align}
    E_\qq = \frac{M_\qq}{N_\qq} 
\end{align}
To illustrate the stiffness of the Goldstone mode, and also its gapless nature, we perform a small $|\qq|$ expansion and find
\begin{align}
   & M_{\qq} \approx 
    \hubbard {A}\frac{1}{N}\sum_{\kk,\mu\nu}
    \bigg[ Q^{\mu\nu}(\kk)
  -\frac{ \partial_\mu \delta \epsilon_{\kk,1} \partial_\nu \delta \epsilon_{\kk,1} + 
    \hubbard {A} \partial_\mu \partial_\nu \delta \epsilon_{\kk,1}
    }{({A}\hubbard )^2}
    \bigg] q^\mu q^\nu \nonumber\\ 
    &N_{\qq} \approx  1 +  O(|\qq|^2)
\end{align}
where 
\begin{align}
    Q^{\mu\nu}(\kk) = \sum_{aa'} \partial_{\nu}U^*_{\kk,a'1}
    \bigg[ \delta_{a,a'} - U_{\kk,a'1}U_{\kk,a1}^*
    \bigg] 
   \partial_\mu U_{\kk,a1} 
\end{align} 
We now show that for an isolated band, 
\begin{align}
    \sum_{\kk} \partial_\mu \partial_\nu \delta \epsilon_{\kk,1} =0 
\end{align}
For an isolated narrow band, its dispersion $\delta \epsilon_{\kk,1}$ is a smooth and periodic function of $\kk$. This allows us to rewrite $\delta \epsilon_{\kk,1}$ as 
\begin{align}
   \delta \epsilon_{\kk,1} = \sum_\RR T(\RR)e^{i\kk\cdot\RR}
\end{align}
with $T(\RR)$ the Fourier transformation of $\delta \epsilon_{\kk,1}$. 

This indicates
\begin{align}
 \sum_{\kk} \partial_\mu \partial_\nu \delta \epsilon_{\kk,1} = \sum_\kk \sum_\RR T(\RR)(-R^\mu R^\nu) e^{i\kk\cdot\RR} = \sum_\RR T(\RR) (-R^\mu R^\nu) N\delta_{\RR,0}  =0    
\end{align}
Therefore, we have 
\ba 
 & M_{\qq} \approx 
    \hubbard {A}\frac{1}{N}\sum_{\kk,\mu\nu}
    \bigg[ Q^{\mu\nu}(\kk)
  -\frac{ \partial_\mu \delta \epsilon_{\kk,1} \partial_\nu \delta \epsilon_{\kk,1} + 
    \hubbard {A} \partial_\mu \partial_\nu \delta \epsilon_{\kk,1}
    }{({A}\hubbard )^2}
    \bigg] q^\mu q^\nu 
    = \hubbard {A}\frac{1}{N}\sum_{\kk,\mu\nu}
    \bigg[ Q^{\mu\nu}(\kk)
  -\frac{ \partial_\mu \delta \epsilon_{\kk,1} \partial_\nu \delta \epsilon_{\kk,1}
    }{({A}\hubbard )^2}
    \bigg] q^\mu q^\nu 
\ea 
Therefore we conclude the dispersion of the Goldstone mode is 
\begin{align}
    E_\qq =\frac{M_\qq}{N_\qq} \approx  \hubbard {A} \frac{1}{N}\sum_{\kk,\mu\nu}\bigg[Q^{\mu\nu}(\kk) - \frac{\partial_\mu \delta \epsilon_{\kk,1} \partial_\nu \delta \epsilon_{\kk,1}}{({A}\hubbard )^2}\bigg] q^\mu q^\nu 
\end{align}
We let 
\begin{align}
    Q^{\mu\nu}  =\frac{1}{N}\sum_\kk Q^{\mu\nu}(\kk),\quad M^{\mu\nu} = \frac{1}{N}\sum_\kk \partial_\mu \delta \epsilon_{\kk,1}\partial_\nu \delta \epsilon_{\kk,1}
\end{align}
where $Q^{\mu\nu}$ denotes the quantum geometry of the system, and $M^{\mu\nu}$ characterizes the dispersion of the narrow band. Then we find 
\begin{align}
\label{eq:stiffness_ferro_small_q}
    E_\qq \approx \hubbard {A} \sum_{\mu\nu} \bigg[Q^{\mu\nu} - \frac{M^{\mu\nu}}{({A}\hubbard )^2}\bigg]q^\mu q^\nu 
\end{align}
We find that the dispersion of the Goldstone mode is quadratic. We now show it is consistent with the counting rule discussed in Refs.\cite{nielsen1976count,watanabe2012unified}. 
The original system possesses an SU(2) spin symmetry with three generators. After the development of ferromagnetic order, this symmetry is spontaneously broken down to a U(1) spin symmetry with only one generator. Since the translational symmetry remains unbroken, the number of spontaneously broken generators is $n_{{BG}} = 2$. 
According to the counting rule of the Goldstone modes~\cite{nielsen1976count,watanabe2012unified}, the number of Goldstone modes must satisfy the inequality 
\begin{equation}
\label{eq:NC_inequality}
n_I + 2n_{II} \ge n_{\mathrm{BG}},
\end{equation}
where $n_I$ is the number of type-I Goldstone modes with dispersion proportional to odd powers of momentum, and $n_{II}$ is the number of type-II Goldstone modes with dispersion proportional to even powers of momentum. 
From our Holstein–Primakoff boson analysis, we find that there is only one Goldstone mode. Therefore, in order for \cref{eq:NC_inequality} to be satisfied, this mode must be type-II. Consequently, its dispersion must be proportional to even powers of momentum. Indeed, our calculation confirms that the Goldstone mode exhibits a quadratic dispersion.

We observe that the quantum geometry tends to stabilize the ferromagnetic state, while the dispersions of the narrow bands tend to destabilize it. The instability of ferromagnetic states is characterized by the parameter values where $Q^{\mu\nu} - \frac{M^{\mu\nu}}{(A\hubbard )^2}$ has a negative eigenvalue. 
However, the transition between antiferromagnetism and ferromagnetism estimated from stiffness (\cref{eq:stiffness_ferro_small_q}) is different from the transition point estimated from the effective spin-spin coupling (\cref{eq:Eq_expansion_small_q}). 
Specifically, we consider the case $Q^{\mu\nu} = Q \delta_{\mu\nu}$ and $M^{\mu\nu} = M \delta_{\mu\nu}$. From the stiffness calculation in \cref{eq:stiffness_ferro_small_q}, the transition point is $Q = M/(A\hubbard)^2$, whereas the effective spin model in \cref{eq:Eq_expansion_small_q} yields $Q = 2M/(A^2\hubbard)^2$.
The deviation comes from the fact that while computing spin stiffness, the high-order spin-spin interaction terms are also included.  
This can be seen from the perfect flat-band limit, where the stiffness is the velocity of the Goldstone mode and can be obtained exactly from the expansion we have performed in \cref{eq:stiffness_ferro_small_q}. However, while we are calculating the spin-spin couplings, we truncated to the two-body interactions. The higher-order interactions, such as interactions that take the form of 
$n^\mu_{x_i}n^\nu_{x_j}n^\delta_{x_m}n^\gamma_{x_n}$ are not included. 
This high-order term also contributes to the spin stiffness and has been implicitly included in our calculation of \cref{eq:stiffness_ferro_small_q} discussed in this section, yielding the exact result in the perfect flat-band limit. 

Finally, we note that although our effective spin-spin couplings neglect higher-order terms, they still capture essential information about the magnetic orderings favored by the wavefunctions and dispersions of the system.

\section{Toy model}
\label{sec:toy_model}
To test our theory, we discuss the magnetism of a simple toy model using both our analytic results as well as numerical analysis. We show that a transition between a ferromagnetic phase and an antiferromagnetic phase can be realized by tuning the quantum geometry and the bandwidth of the narrow band. 

We take a system formed by two layers of square lattices, with each atom located at 
\ba 
\{ (n a_0, ma_0,z_0/2) | n,m\in \mathbb{Z}\} \cup 
\{ (n a_0, ma_0,-z_0/2) | n,m\in \mathbb{Z}\} 
\ea 
$a_0$ denotes the lattice constant along $x,y$ directions, $z_0$ denotes the distance between two layers.
$c_{\RR,l, \sigma}^\dag $ creates a electron at site $\RR \in (a_0\mathbb{Z},0) +(0,a_0\mathbb{Z})$ layer $l = \pm $ with spin $\sigma=\up/\dn$. We also introduce the electron operators in the momentum space as 
\ba 
c_{\kk,l,\sigma} = \frac{1}{\sqrt{N}}\sum_\RR c_{\RR,l,\sigma} e^{-i\kk\cdot\RR }
\ea 
We consider the following tight-binding models 
\ba 
H_0 = \sum_{\sigma,\kk}
\begin{bmatrix}
    c_{\kk,+,\sigma}^\dag & c_{\kk,-,\sigma}^\dag 
\end{bmatrix}
\cdot 
\begin{bmatrix}
    \epsilon_\kk -\mu & v e^{i\alpha_\kk } \\
 v e^{-i\alpha_\kk } & \epsilon_\kk-\mu  
\end{bmatrix}\cdot 
\begin{bmatrix}
    c_{\kk,+,\sigma} \\ c_{\kk,-,\sigma}
\end{bmatrix}
\ea 
where 
\ba 
&\epsilon_\kk = -2t (\cos(k_x) + \cos(k_y)) \nonumber\\ 
&\alpha_\kk = \zeta (\cos(k_x)+\cos(k_y))
\ea 
The Hamiltonian is characterized by $t, v, \zeta$. This model is initially introduced in Ref.~\cite{Hofmann2022,PhysRevLett.130.226001}.
In the space, we have 
\ba 
H_0 = 
\sum_{\RR,\Delta\RR,\sigma}
\begin{bmatrix}
    c_{\RR,+,\sigma}^\dag & c_{\RR,-,\sigma}^\dag 
\end{bmatrix}
\cdot 
\begin{bmatrix}
    t_{\Delta\RR} -\mu & t'_{\Delta\RR} 
    % v e^{i\alpha_\kk } 
    \\
 % v e^{-i\alpha_\kk }
 t'^*_{\Delta\RR} & t_{\Delta\RR}-\mu  
\end{bmatrix}\cdot 
\begin{bmatrix}
    c_{\RR+\Delta\RR,+,\sigma} \\ c_{\RR+\Delta\RR,-,\sigma}
\end{bmatrix}
\ea 
We have a nearest-neighbor intra-layer hopping
\ba 
t_{\Delta\RR} = -t \bigg[\delta_{\Delta\RR, (a_0,0)}
+\delta_{\Delta\RR, (-a_0,0)}
+\delta_{\Delta\RR, (0,a_0)}
+\delta_{\Delta\RR, (0,-a_0)}\bigg] 
\ea 
and the following inter-layer hopping 
\ba 
t'_{\Delta\RR = (n_x a_0, n_ya_0)} = \frac{1}{N}\sum_\kk ve^{i\alpha_\kk }e^{-i\kk\cdot\Delta\RR}
= v (i)^{n_x+n_y}\mathcal{J}_{n_x}(\zeta) 
\mathcal{J}_{n_y}(\zeta) 
\ea 
For $n_x \gg 1, n_y \gg 1$, we approximately have 
\ba 
t'_{\Delta\RR = (n_x a_0, n_ya_0)}
\sim v \frac{ (i)^{n_x+n_y}}{2\pi n_xn_y}
(\frac{ e\zeta }{2 n_x})^{n_x}
(\frac{ e\zeta }{2 n_y})^{n_y}
\ea

% The off-diagonal term corresponds to long-range hopping. 
% Both the dispersion and the quantum geometry of the narrow band can be easily stunned in the above model. 
The eigenvalues and eigenstates of the hopping matrix are 
\ba 
&E_{\kk,1/2} = -\mu +\epsilon_\kk \mp v  \nonumber\\ 
&U_{\kk,l1} = \frac{1}{\sqrt{2}}
\begin{bmatrix}
    1 & -e^{-i\alpha_\kk} 
\end{bmatrix}_l,\quad 
U_{\kk,l2} = \frac{1}{\sqrt{2}}
\begin{bmatrix}
    1 & e^{-i\alpha_\kk} 
\end{bmatrix}_l
\ea 
We first observe that the gap between two bands is determined by $v$. 
We can take the limit where $v\gg  t >0$ and let $\mu = -v$. This indicates the lowest-energy band with energy $E_{\kk,1}$ and eigenvector $U_{\kk,l1}$ appears near the Fermi energy. The other band (characterized by $E_{\kk,2},U_{\kk,l2}$) is far away from Fermi energy. 

We first note that, by setting $t=0$, we realize a perfect flat band. Gradually increasing $t$ will gradually increase the bandwidth of the narrow band (lowest-energy band). The bandwidth of the narrow band $D=8t$ is proportional to the hopping $t$. Next, we calculate the quantum geometry of the narrow band. We fine
\ba 
&Q^{\mu\nu}(\kk) = \sum_l \partial_{k^\mu} U_{\kk,l1}^* \partial_{k^\nu} U_{\kk,l1} 
- \sum_l \partial_{k^\mu} U_{\kk,l1}^*  U_{\kk,l1} 
\sum_{l'} \partial_{k^\nu} U_{\kk,l'1}  U_{\kk,l'1} ^* \nonumber\\ 
=& \frac{1}{4}\partial_{k^\mu}\alpha_\kk \partial_{k^\nu}\alpha_\kk =\frac{\zeta^2}{4}\sin(k^\mu) \sin(k^\nu) \nonumber\\ 
&Q^{\mu\nu} = \frac{1}{N}\sum_\kk Q^{\mu\nu}(\kk) = \frac{\zeta^2}{8}\delta_{\mu,\nu},\quad Q = \sum_\mu Q^{\mu\nu} = \zeta^2/4 
\label{eq:quantum_geometry_toy_model}
\ea 
This indicates the quantum geometry can be tuned by  $\zeta$. 
We also note that the quantum geometry we considered here corresponds to the minimal quantum geometry\cite{PhysRevB.102.165148,herzog-arbeitman_many-body_2022,huhtinen_revisiting_2022}. 
Therefore, we can tune the quantum geometry and the dispersion of the narrow bands independently by tuning the $t$ and $\zeta$ respectively. 

We next consider the interaction terms. We consider the on-site Hubbard interaction 
\ba 
H_{int} =\frac{\hubbard }{2} \sum_{\RR,l} \left(\sum_\sigma c_{\RR,l,\sigma}^\dag c_{\RR,l,\sigma}-\frac{1}{2}\right)^2 
\ea 
where the additional $1/2$ comes from the normal ordering of the system with respect to the ground states of the non-interacting Hamiltonian at $\mu =-v$. 

We now solve the model in the limit of 
\ba 
v \gg  \hubbard  \gg  t 
\ea 
where $v\gg \hubbard $ ensures only the narrow band near the Fermi energy contributes to the low-energy physics. $\hubbard \gg t$ indicates the interaction is much larger than the bandwidth of the narrow band near the Fermi energy. 

We now derive effective spin-spin interactions. We first calculate $A_{(\RR,l), (\RR',l')}$
\ba 
&A_{(\RR,l),(\RR',l)} = \frac{1}{2} \delta_{\RR,\RR'} \nonumber\\ 
&A_{(\RR,+),(\RR',-)} = \frac{1}{N}\sum_\kk \frac{1}{2}e^{i\kk\cdot(\RR-\RR')}(-1)e^{i\alpha_\kk} 
\label{eq:toy_model_A}
\ea 
To evaluate the momentum summation, we transform the summation to an integral and use the following expansions 
\ba 
\sum_{n=-\infty}^{\infty}(i)^n \mathcal{J}_n(\zeta) e^{in k} = e^{i\zeta \cos(k)} 
\ea 
where $\mathcal{J}_n(\zeta)$ is the Bessel function of the first kind. 

We then have 
\ba 
&A_{(\RR,+),(\RR',-)} = \frac{-1}{8\pi^2} \left[
\int_{-\pi}^\pi \sum_{n=-\infty}^{\infty}(i)^n \mathcal{J}_n(\zeta) e^{in k}e^{i k_x (R_x-R'_x)/a_0} dk_x  \right] 
 \left[
\int_{-\pi}^\pi \sum_{n=-\infty}^{\infty}(i)^n \mathcal{J}_n(\zeta) e^{in k}e^{i k_y (R_y-R'_y)/a_0} dk_y  \right] \\ 
=&\frac{-i^{-(R_x-R_x')/a_0 -(R_y-R_y')/a_0}}{2}\mathcal{J}_{(R_x'-R_x)/a_0}(\zeta) 
\mathcal{J}_{(R_y'-R_y)/a_0}(\zeta) \, . 
\label{eq:toy_model_A_2}
\ea 

% At large distances with $|\RR-\RR'| >> \zeta$, we have 
% \ba 
% A_{(\RR,+),(\RR',-)} 
% \approx &\frac{-1}{2}\left(-i\zeta/2\right)^{ |R_x-R_x'|/a_0+|R_y-R_y'|/a_0}
% \ea 
In addition, we note that 
\ba 
A_{(\RR,-),(\RR',+)} = (A_{(\RR',+),(\RR,-)})^*
\ea 

We next investigate 
\ba 
B_{(\RR,l),(\RR',l)} =& \frac{1}{N}\sum_\kk \epsilon_\kk  \frac{1}{2} e^{i\kk\cdot(\RR-\RR')} = -\frac{t}{2} \left(\delta_{R_x-R_x',a_0}+\delta_{R_x-R_x',-a_0}+\delta_{R_y-R_y',a_0}+\delta_{R_y-R_y',-a_0} \right) \nonumber\\ 
B_{(\RR,+),(\RR',-)} = &\frac{1}{N}\sum_\kk \epsilon_\kk  \frac{-1}{2}e^{i\alpha_\kk} e^{i\kk\cdot(\RR-\RR')} \nonumber\\ 
=&    t i^{-(R_x-R_x')/a_0-(R_y-R_y')/a_0-1} 
\left[ \mathcal{J}'_{(R_x'-R_x)/a_0}(\zeta) \mathcal{J}_{(R_y'-R_y)/a_0}(\zeta) 
+\mathcal{J}_{(R_x'-R_x)/a_0}(\zeta) \mathcal{J}'_{(R_y'-R_y)/a_0}(\zeta) 
\right] 
\nonumber\\ 
B_{(\RR,-),(\RR',+)} = &
B_{(\RR',+),(\RR,-)}^* \nonumber\\
B = & 0 
\label{eq:toy_model_B}
\ea 

Finally, we note that 
\ba 
C_{(\RR,l),(\RR',l)} =& \frac{1}{N}\sum_\kk \epsilon_\kk^2 \frac{1}{2} e^{i\kk\cdot(\RR-\RR')} = \frac{t^2}{2}\sum_{\bm{e}_1,\bm{e}_2\in \{(1,0),(0,1),(-1,0),(0,-1)\}}  \delta_{\RR-\RR', \bm{e}_1+\bm{e}_2}  \nonumber\\ 
C_{(\RR,+),(\RR',-)} =&\frac{1}{N}\sum_\kk (\delta \epsilon_\kk)^2 \frac{-e^{i\alpha_\kk}}{2} e^{i\kk\cdot(\RR-\RR')}
 \nonumber\\ 
=&-\frac{t^2}{2}\sum_{\bm{e}_1,\bm{e}_2 \in \{(a_0,0),(-a_0,0),(0,a_0),(0,-a_0) \} } 
i^{-(R_x-R_x'+e_{1,x}+e_{2,x})/a_0 -(R_y-R_y'+e_{1,y}+e_{2,y})/a_0 }\nonumber\\ 
&
\mathcal{J}_{-(R_x-R_x'+e_{1,x}+e_{2,x})/a_0}(\zeta) 
\mathcal{J}_{-(R_y-R_y'+e_{1,y}+e_{2,y})/a_0}(\zeta) 
\nonumber\\ 
% -\frac{t^2}{2} i^{-(R_x-R_x'+R_y-R_y')/a_0  }
% \left[ 
% -\left( J_{-(R_x-R_x')/a_0-2}(\zeta)
% +J_{-(R_x-R_x')/a_0+2}(\zeta)\right) J_{-(R_y-R_y')/a_0}(\zeta) \right. \nonumber\\ 
% &\left. 
% -\left( J_{-(R_y-R_y')/a_0-2}(\zeta)
% +J_{-(R_y-R_y')/a_0+2}(\zeta)\right) J_{-(R_x-R_x')/a_0}(\zeta) \right. \nonumber\\ 
% &
% + 2 \left[2 J_{-(R_x-R_x')/a_0}(\zeta)  J_{-(R_y-R_y')/a_0}(\zeta)\right. \nonumber\\ 
% &\left.   
% +J_{-(R_x-R_x')/a_0+1}(\zeta)  J_{-(R_y-R_y')/a_0-1}(\zeta)
% +J_{-(R_x-R_x')/a_0-1}(\zeta)  J_{-(R_y-R_y')/a_0+1}(\zeta)
% \right] \nonumber\\ 
% & \left. 
% -J_{-(R_x-R_x')/a_0+2}(\zeta)  J_{-(R_y-R_y')/a_0}(\zeta)
% -J_{-(R_x-R_x')/a_0-2}(\zeta)  J_{-(R_y-R_y')/a_0}(\zeta)\right. \nonumber\\ 
% & \left. 
% -J_{-(R_x-R_x')/a_0}(\zeta)  J_{-(R_y-R_y')/a_0+2}(\zeta)
% -J_{-(R_x-R_x')/a_0}(\zeta)  J_{-(R_y-R_y')/a_0-2}(\zeta)
% \right]\nonumber\\ 
C_{(\RR,-),(\RR',+)} = &C_{(\RR',+),(\RR,-)}^*
\nonumber\\
C =&2 t^2  
\label{eq:toy_model_C}
\ea

We the use Eq.~\ref{eq:def_parameters}, Eq.~\ref{eq:toy_model_A}, Eq.~\ref{eq:toy_model_A_2}, Eq.~\ref{eq:toy_model_B}, and Eq.~\ref{eq:toy_model_C}, we find 
\ba 
J_{(\RR,l),(\RR',l)} =& \frac{t^2}{4\hubbard  A^4} \left(\delta_{R_x-R_x',a_0}+\delta_{R_x-R_x',-a_0}+\delta_{R_y-R_y',a_0}+\delta_{R_y-R_y',-a_0} \right)  \nonumber\\ 
J_{(\RR,+),(\RR',-)} = & -
\bigg( \frac{\hubbard }{16}
+ \frac{3t^2}{2A^5\hubbard}
\bigg) 
\left[ \mathcal{J}_{R_x'-R_x}(\zeta) \mathcal{J}_{R_y'-R_y}(\zeta)\right]^2  \nonumber\\ 
% &+ \frac{-2t^2}{4A^4\hubbard }\sum_{\bm{e}\in \{(0,1),(0,-1),(1,0),(-1,0)\}}J_{-(R_x-R_x')}(\zeta) 
% J_{-(R_y-R_y')}(\zeta) 
% J_{-(R_x-R_x'-2e_x)}(\zeta) 
% J_{-(R_y-R_y'-2e_y)}(\zeta)  \nonumber\\ 
% &+\frac{2t^2}{A^4\hubbard }[J_{-(R_x-R_x')}(\zeta) 
% J_{-(R_y-R_y')}(\zeta) ]^2 \nonumber\\ 
&+ \frac{t^2}{2A^4\hubbard }\sum_{\bm{e}_1,\bm{e}_2 \in \{(a_0,0),(-a_0,0),(0,a_0),(0,-a_0) \} }
\mathcal{J}_{-(R_x-R_x')/a_0}(\zeta) 
\mathcal{J}_{-(R_y-R_y')/a_0}(\zeta)  \nonumber\\ 
&
\mathcal{J}_{-(R_x-R_x'-e_{1,x}-e_{2,x})/a_0}(\zeta) 
\mathcal{J}_{-(R_x-R_x'-e_{1,y}-e_{2,y})/a_0}(\zeta) 
\nonumber\\ 
&+ \frac{t^2}{A^4 \hubbard }\left(\mathcal{J}_{-(R_y-R_y')}(\zeta) \mathcal{J}_{-(R_x-R_x')}'(\zeta) +  \mathcal{J}_{-(R_x-R_x')}(\zeta) \mathcal{J}_{-(R_y-R_y')}'(\zeta)
\right) ^2 \nonumber\\ 
J_{(\RR,-),(\RR',+)} = &J_{(\RR',+),(\RR,-)}
\ea 
In addition, the parameter $A$ is introduced in Eq.~\ref{eq:def_para_A} and takes the value of 
\ba
 A =  \frac{1}{2} 
\ea

We now discuss the spin interaction terms. We first investigate the ``on-site'' coupling between electrons of two layers. Here, ``on-site'' indicates the electrons have the same $x,y$ coordinates. The corresponding coupling is 
\ba 
J_{(\RR,+),(\RR,-)} = -\bigg( \frac{\hubbard }{16}
+ \frac{3t^2}{2A^5\hubbard}\bigg) [\mathcal{J}_0(\zeta)]^4  +\frac{32t^2\mathcal{J}_{0}(\zeta)^2}{\hubbard  }\bigg( \mathcal{J}_{0}(\zeta)\mathcal{J}_{2}(\zeta) + [\mathcal{J}_{1}(\zeta)]^2  
\bigg) 
\ea 
We can assume both quantum geometry and dispersion are weak. To be more specific, we assume $\zeta^2, t^2/\hubbard ^2$ are small and are in the same order. Then we could perform an expansion in both $\zeta^2, t^2/\hubbard ^2$ which gives 
\ba 
J_{(\RR,+),(\RR,-)} \approx
-\frac{\hubbard }{16} (1-\zeta^2) 
-\frac{3t^2}{2A^5\hubbard}
+O(\zeta^4, t^4/\hubbard ^4,t^2/\hubbard ^2\zeta^2) 
\ea 
This gives a dominant on-site ferromagnetic coupling between electrons of two layers. 

The nearest-neighbor intra-layer coupling is 
\ba 
J_{(\RR,l),(\RR+\bm{e},l)} = \frac{4t^2}{\hubbard },\quad \bm{e} \in \{(a_0,0),(-a_0,0),(0,a_0),(0,-a_0) \}
\ea 
which is anti-ferromagnetic. 
The nearest-neighbor inter-layer coupling is 
\ba 
J_{(\RR,l),(\RR+\bm{e},-l)} =& -\bigg( \frac{\hubbard }{16}
+ \frac{3t^2}{2A^5\hubbard}\bigg) [\mathcal{J}_0(\zeta) \mathcal{J}_{1}(\zeta)]^2 \nonumber\\
&
+\frac{16t^2}{\hubbard \zeta^2} 
\bigg[  4\zeta \mathcal{J}_{0}(\zeta) \mathcal{J}_1(\zeta)^3 + \zeta^2 \mathcal{J}_{1}(\zeta)^4 + \mathcal{J}_{0}(\zeta)^2 (1-4\zeta^2)\mathcal{J}_{1}(\zeta)^2 + \mathcal{J}_{0}(\zeta)^2 \zeta^2 \mathcal{J}_{2}(\zeta)^2 
\bigg] \nonumber\\ 
\bm{e} \in  &\{(a_0,0),(-a_0,0),(0,a_0),(0,-a_0) \}
\ea 
We perform an expansion in both $\zeta^2, t^2/\hubbard ^2$ which gives
\ba 
J_{(\RR,l),(\RR+\bm{e},-l)} \approx \hubbard \bigg[\frac{4t^2}{\hubbard ^2}-\frac{\zeta^2\hubbard }{64}\bigg] +O(\zeta^4, t^4/\hubbard ^4,t^2/\hubbard ^2\zeta^2) 
\ea 

Therefore, by truncating to the nearest-neighbor couplings, the effective spin model can be approximately written as  
\ba 
H = &-\bigg[ 
\frac{\hubbard }{16}(1-\zeta^2)
+\frac{3t^2}{2A^5\hubbard}\bigg]
\sum_{\RR,l} \bm{n}_{\RR,l}\cdot \bm{n}_{\RR,-l} \nonumber\\
&+  \sum_{\RR, \bm{e} \in\{(1,0),(-1,0),(0,1),(0,-1) \} , l}\bigg[ \frac{4t^2}{\hubbard } \bm{n}_{\RR,l}\cdot \bm{n}_{\RR,l} +(
- \frac{\hubbard \zeta^2 }{64} + \frac{4t^2}{\hubbard })  \bm{n}_{\RR,l}\cdot \bm{n}_{\RR,-l} 
\bigg] 
\ea 
% We further assume $t^2/\hubbard ^2 \sim \zeta^2 $ which are much smaller than 1. 
% Then, we find 
% \ba 
% H = -\frac{\hubbard }{16}(1-\zeta^2) \sum_{\RR,l} \bm{n}_{\RR,l}\cdot \bm{n}_{\RR,-l}  +  \sum_{\RR, \bm{e} \in\{(1,0),(-1,0),(0,1),(0,-1) \} , l}\bigg[ \frac{4t^2}{\hubbard } \bm{n}_{\RR,l}\cdot \bm{n}_{\RR+\bm{e},l} 
% - \frac{\hubbard \zeta^2 }{64}   \bm{n}_{\RR,l}\cdot \bm{n}_{\RR+\bm{e},-l} 
% \bigg] 
% \ea 
We can now discuss the ground state of the effective spin model. First, the on-site inter-layer ferromagnetic coupling is dominant which always align the spin of electrons from two layers with the same $x,y$ coordinates. Then we can introduce the effective spin operators 
\ba 
\bm{n}_\RR = \bm{n}_{\RR,+}+\bm{n}_{\RR,-}
\ea 
which is the sum of two spin operators of two layers. 
Then the effective couplings, projecting to $n_\RR$ spins, are
\ba 
 H_{eff}=  \sum_{\RR, \bm{e} \in\{(1,0),(-1,0),(0,1),(0,-1) \} , l}\hubbard \bigg[ \frac{4t^2}{\hubbard ^2}
- \frac{\zeta^2 }{128}  
\bigg]  \bm{n}_{\RR}\cdot \bm{n}_{\RR+\bm{e}} 
\ea 

We can observe, that the system tends to develop a ferromagnetic order when the quantum geometry dominates ($\frac{\zeta^2 }{128}>\frac{4t^2}{\hubbard ^2}
$),  since the interaction is ferromagnetic. 
However, when the hopping contribution dominates ($\frac{4t^2}{\hubbard ^2}
>\frac{\zeta^2 }{128}$), the coupling becomes antiferromagnetic which favors an antiferromagnetic order. 
Then a transition between the ferromagnetic phase and the antiferromagnetic phase happens at 
\ba 
4t^2/\hubbard  = \hubbard \zeta^2 /128 
% \Rightarrow  Q = \frac{128t^2}{\hubbard ^2}
\ea 
Using $Q = \zeta^2/4, M =4t^2$ (from Eq.~\ref{eq:def_mass_Q_matrix}, Eq.~\ref{eq:quantum_geometry_toy_model}),
 we can find the transition point is 
\ba 
Q = \frac{32M}{\hubbard ^2 } = \frac{2M}{A^4\hubbard ^2}
\label{eq:phase_boundary_ana_toy_model}
\ea 
which also matches our previous result in Eq.~\ref{eq:Eq_expansion_small_q}.

To confirm our analytical result, we also perform a Hartree-Fock study by taking $\hubbard =10$ and $v=50$. We treat $\zeta, t$ as our tuning parameters. The Hartree-Fock phase diagram of the model has been shown in Fig.~\ref{fig:pd_toy_model}, where we can observe the analytical phase boundary matches well with the numerical phase boundary at small $\zeta, |t|$. As we gradually increase $\zeta, |t|$, the high order terms (terms at the order of $|\zeta|^4 , t^4/\hubbard ^4, |\zeta|^2 t^2/\hubbard ^2$) may make more and more contributions, and our estimated phase boundary in Eq.~\ref{eq:phase_boundary_ana_toy_model} becomes less reliable.

\begin{figure}
    \centering
    \includegraphics[width=0.5\linewidth]{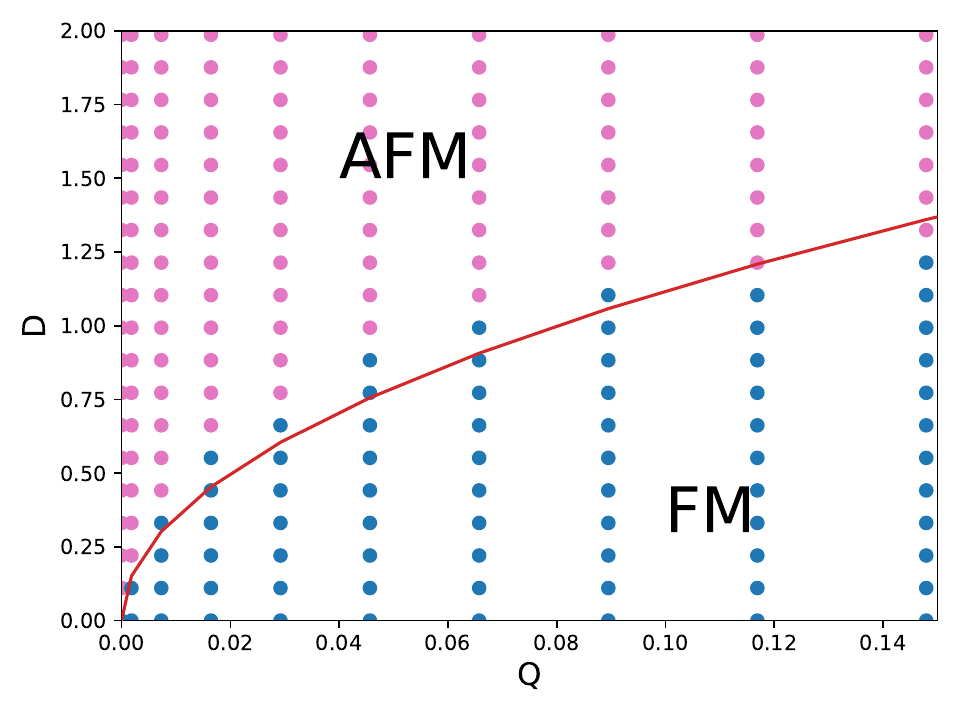}
    \caption{Hartree Fock phase diagram. $Q$ denotes the quantum geometry of the narrow band and $D =8t$ denotes the bandwidth of the narrow band. Pink and blue denote the ground state from the Hartree-Fock calculations, where pink denotes the antiferromagnetic phase and blue denotes the ferromagnetic phase. The red curve denotes the approximated phase boundary we obtained from the analytical calculations (Eq.~\ref{eq:phase_boundary_ana_toy_model}). }
    \label{fig:pd_toy_model}
\end{figure}

\section{Matsubara summations}
\label{sec:Matsubara_summation}
We aim to evaluate the following summations 
\ba 
\frac{1}{\beta} \sum_{\omega=(2n+1)\pi/\beta}  \frac{(i\omega)^n}{(\omega^2 +\omega_0^2)^2}e^{i\omega 0^+}
\ea 
for an integer number $n\ge 0$ and a real number $\omega_0>0$. 

We define 
\ba 
f(z) =  \frac{z^n}{(-z^2 +\omega_0^2)^2} \frac{1}{z^n} e^{z 0^+}
\ea 
We could use the contour integral which indicates 
\ba 
\frac{1}{\beta}\sum_{i\omega} f(i\omega) = \frac{1}{2\pi i\beta }
\oint f(z) h(z) dz  = -\frac{1}{\beta} \sum_{z_0 \in Poles}\text{Res}[f(z_0)]h(z_0)
\ea 
where 
\ba 
h(z) = - \beta n_F(z) 
\ea 

$f(z)$ has two poles at $\omega_0,-\omega_0$. The corresponding residues are 
\ba 
&\text{Res}[f(\omega_0) ]  =\frac{d}{dz}[(z-\omega_0)^2f(z)]\bigg|_{z = \omega_0}=  \frac{n-1}{4}\omega_0^{n-3} 
\nonumber\\ 
&\text{Res}[f(-\omega_0) ]  = \frac{d}{dz}[(z+\omega_0)^2f(z)]\bigg|_{z = -\omega_0} =\frac{n-1}{4}(-\omega_0)^{n-3} 
\ea 

In the low-temperature limit, we have 
\ba 
h(\omega_0) \approx 0 ,\quad h(-\omega_0)\approx-\beta 
\ea
Then 
\ba 
\frac{1}{\beta} \sum_{\omega=(2n+1)\pi/\beta}  \frac{(i\omega)^n}{(\omega^2 +\omega_0^2)^2}e^{i\omega 0^+}= \frac{n-1}{4}(-\omega_0)^{n-3}
\label{eq:mats_sum}
\ea 

\end{document}